\documentclass[aps,pre,twocolumn]{revtex4-1} 	

\usepackage{graphicx}

\usepackage{amssymb,amsfonts,amsmath}
\usepackage{hyperref} 
\hypersetup{
    colorlinks,
    citecolor=blue,
    filecolor=black,
    linkcolor=red,
    urlcolor=blue
}

\def \>{\rangle} 
\def \<{\langle} 
 
\def\be{\begin{equation}} 
\def\ee{\end{equation}} 
\def\longrightharpoonup{\relbar\joinrel\rightharpoonup}
\def\longleftharpoondown{\leftharpoondown\joinrel\relbar}

\def\longrightleftharpoons{
  \mathop{
    \vcenter{
      \hbox{
      \ooalign{
        \raise1pt\hbox{$\longrightharpoonup\joinrel$}\crcr
	  \lower1pt\hbox{$\longleftharpoondown\joinrel$}
	  }
      }
    }
  }
}

\newcommand \bea {\begin{eqnarray}} 
\newcommand \eea {\end{eqnarray}}

\usepackage{xcolor}
\usepackage{physics}

\begin{document}

\title{The Perturbative Resolvent Method: spectral densities of random matrix ensembles via perturbation theory}

\author{Wenping Cui}
\email{cuiw@bu.edu}
\affiliation{Dept. of Physics, Boston University, Boston, MA 02215}
\affiliation{Dept. of Physics, Boston College, Chestnut Hill, Mass. 02467}

\author{Jason W. Rocks}
\email{jrocks@bu.edu}
\author{Pankaj Mehta}
\email{pankajm@bu.edu}
\affiliation{Dept. of Physics, Boston University, Boston, MA 02215}
\affiliation{Faculty of Computing and Data Sciences, Boston University, Boston, MA 02215}

\begin{abstract}
We present a simple, perturbative approach for calculating spectral densities for random matrix ensembles
in the thermodynamic limit we call the Perturbative Resolvent Method (PRM). The PRM is based on constructing 
a linear system of equations and calculating how the solutions to these equation 
change in response to a small perturbation using the zero-temperature cavity method. We illustrate the power of the method
by providing simple analytic derivations of  the Wigner Semi-circle Law for symmetric matrices, the Marchenko-Pastur Law for Wishart matrices, the spectral density for a product Wishart matrix composed of two square matrices, and the Circle and elliptic laws
for real random matrices. 
\end{abstract}
\maketitle

Random matrices are central to variety of problems ranging from statistical physics \cite{dyson1962statistical, auffinger2013random}  to  quantum chaos \cite{kriecherbauer2001random}, ecology \cite{may1972will, allesina2015stability, biroli2018marginally}, and wireless communication \cite{couillet2011random}.
An important problem in Random Matrix Theory (RMT) is to calculate the spectral density of an ensemble of random matrices \cite{livan2018introduction}. The spectral density measures the density of eigenvalues in the complex plane and plays a central role in many RMT-based approaches. There exist numerous methods for calculating 
spectral densities, including many prominent methods that draw heavily from the physics of disordered systems such as the replica method \cite{kuhn2008spectra}, large-N diagrammatic expansions \cite{sengupta1999distributions}, and the finite temperature cavity method \cite{rogers2008cavity, rogers2009cavity}. 

Here, we introduce a new simple, flexible, perturbative approach for calculating spectral densities in the thermodynamic limit (i.e. in the limit where the size of the matrices become infinitely large). Our method is inspired by recent zero temperature cavity calculations in the context of ecology \cite{advani2018statistical, mehta2018constrained, cui2020effect, cui2019diverse}. The central observation underlying the method is the observation that the spectral density can be calculated by constructing an appropriately chosen  system of random linear equations and then asking how the solutions to these equations change in response to small constant perturbations. In particular, we show that the trace of the susceptibility matrix which measures responses to perturbations is precisely the resolvent or Green's function and can be calculated easily using the zero-temperature cavity method.  For this reason, we refer to this approach as the Perturbative Resolvent Method (PRM). Since at its core the PRM is just simply a perturbative way of computing Green's functions, we can make use of many powerful results in RMT and statistical physics relating Green's functions to spectral densities. These include generalizations of the Green's function method to non-Hermitian matrices through hermitian reduction \cite{feinberg1997non, feinberg1997non2}, which associates with each $N$-dimensional non-hermitean ensemble an auxiliary ensemble of  $2N$-dimensional Hermitian matrices. In the PRM, this ``doubling'' of the degrees of freedom simply corresponds to ``doubling'' the the number of random equations, allowing us to also easily calculate spectral densities of many non-Hermitian ensembles.

The PRM assumes a Replica Symmetric Ansatz and exploits the zero-temperature cavity method to calculate Green's functions \cite{brezin1995universal}. An important technical consideration
that makes the method particularly simple to implement is that,  for many random matrix ensembles, there is no need to explicitly solve the resulting self-consistent mean-field cavity equations. Instead, the problem often reduces to simply solving a polynomial equation for the susceptibility. We illustrate this procedure below by giving simple derivations of spectral densities for a number of random matrix ensembles

The paper is organized as follows. We begin by giving some background on the resolvent/Green's function method for Hermitian and non-Hermitian matrices. We then introduce the Perturbative Resolvent Method (PRM) in the context of real symmetric random matrices and show how to derive the Wigner's semi-circle law \cite{wigner1993characteristic}. We then show how this construction can be generalized to calculate the Marchenko-Pastur Law for Wishart matrices  \cite{marchenko1967distribution} and to calculate the spectral density for a product Wishart matrix composed of two square matrices. Finally, we show how the PRM can be generalized to real non-symmetric matrices and provide simple derivations of Girko's circle law \cite{girko1985circular} and elliptic laws \cite{girko1986elliptic} for real random matrices .

\section{Resolvent Methods for Hermitian and Non-Hermitian Matrices}

We begin by briefly summarizing the mathematical results we make use of in the paper. For a full discussion, of the Resolvent/Green's function method in RMT we urge the reader to consult one of the many excellent reviews or textbooks in the field \cite{rogers2010new, livan2018introduction, tao2012topics, couillet2011random,  mingo2017free, bai2010spectral}.

\subsection{Hermitian Matrices}

Let $A$ be a $N \times N$ symmetric matrix. The resolvent or Green's function of $A$ is given by the expression
\be
G_A(z) = {1 \over zI- A}.
\label{Eq:ResolventDef}
\ee
where $z$ is a scalar constant and $I$ is the $N\times N$ identity matrix.
We define the spectral density of $A$ as
\be
\rho_A(x) = {1 \over N} \sum_i \delta(x-\lambda_i),
\ee
where $\lambda_i$ are the eigenvalues of the matrix $A$.  It what follows we will be concerned almost exclusively with the thermodynamic limit 
where $N \rightarrow \infty$ and the spectrum becomes continuous. The spectral density $\rho_A$ of the eigenvalues of $A$ can be 
extracted from the Green's function using the standard relationship
\be
\rho_A(x) = \lim_{\epsilon \rightarrow 0^+}  {1 \over \pi} \mathrm{Im} \qty[{1 \over N}\mathrm{Tr} \, G_A(x-i \epsilon)],
\label{Eq:ResolventTrick}
\ee
where $\mathrm{Tr}$ is the trace.

\subsection{Non-Hermitian Matrices}

Zee and Feinberg \cite{feinberg1997non} generalized this to non-Hermitian matrices (which we also denote by $A$) by carefully considering differentials in the complex plane. 
Generically, the eigenvalues $\lambda_i$ of a non-hermitian matrix are complex and have non-zero real and imaginary parts. The spectral density of $A$ 
over the complex plane $z=x+iy$ is given by
\be
\rho_A(x,y) = {1 \over N} \sum_{i} \delta(x- \mathrm{Re}[ \lambda_i]) \delta(y- \mathrm{Im}[\lambda_i]).
\ee
Again, we will focus on the thermodynamic limit where $N \rightarrow \infty$. 

Zee and Feinberg exploited the fact that over the complex plane on a single Riemann cut
\be
\partial_z  \frac{1}{z^*} = \pi \delta(x)\delta(y)
\label{Eq:CAI}
\ee
to generalize the resolvent formalism above to calculate such spectral functions. 
In particular, they considered a ``symmetrization'' of the matrix $A$ by defining a new  $2N \times 2N$ Hermitian matrix
\be
H(z) = \mqty(
0 & A-z \\
A^*-z^* & 0)
\label{Eq:ZeeH_uncorrected}
\ee
and a corresponding Green's function in the $2N$ dimensional space
\be
\mathcal{G}_A(\eta) = {1 \over \eta I- H}
\label{Eq:GreenComplex}
\ee
where $\eta$ is a scalar constant.
This Green's function can be rewritten in block diagonal form (with each block matrix is of size $N \times N$) as
\be
\mathcal{G}_A(\eta)= 
\mqty(\chi & \nu\\
\nu^* & \chi),
\ee
where in writing this in terms of $\chi$ and $\nu$  we have exploited the symmetries and block diagonal structure of $H(z)$. 
In terms of block matrices, these relationships can be written as
\be
 \mqty(
\eta & zI-A \\
z^*I-A^* & 0)
\mqty(\chi & \nu\\
\nu^* & \chi)
= \mqty(
I & 0 \\
0 & I).
\ee
Focusing on the upper left block, one  has $\eta \chi + ( zI-A)\nu^* =I$,  which for the special case $\eta=0$ implies that the resolvent (Eq. \ref{Eq:GreenComplex}) is 
\be
G_A(z) = {1 \over zI- A} = \nu^*
\ee
and
\be
\rho_A(x,y) ={1 \over \pi} \partial_{z^*}\eval{\qty[ {1 \over N}\Tr \, \nu^*(z, z^*)]}_{\eta=0},
\label{Eq:density-complex_old}
\ee
where in going to the second equation we make use of Eq. \ref{Eq:CAI}.

Here, we slightly modify this prescription. The reason is that a careful analysis actually shows that this formalism, while essentially correct, results
in choosing the wrong complement of the full Riemann sphere for the domain where the spectral density is non-zero (though the boundary between regions
is correctly identified). We note that this mistake is already present in the explicit expressions for the Circle law derived in  \cite{feinberg1997non}. To correct this mistake, we modify a sign in the original construction and consider the $2N$-dimensional non-symmetric matrix
\be
H(z) = \mqty(
0 & A-z \\
-A^*+z^* & 0)
\label{Eq:ZeeH}
\ee 
and once again calculate the corresponding Green's function in the $2N$ dimensional space
\be
\mathcal{G}_A(\eta) = {1 \over \eta I- H}.
\label{Eq:GreenComplex}
\ee
This Green's function can be rewritten in block diagonal form (with each block matrix is of size $N \times N$) as
\be
\mathcal{G}_A(\eta)= \mqty(\chi & \nu\\
-\nu^* & \chi),
\ee
and an almost identical calculation yields that
\be
\rho_A(x,y) =-{1 \over \pi} \partial_{z^*}\eval{\qty[ {1 \over N}\Tr \, \nu^*(z, z^*)]}_{\eta=0}
\label{Eq:density-complex}
\ee
This sign change ensures that the domain where the spectral density is non-zero density occurs on the correct complement of the full Riemann sphere.

\section{The Wigner Semi-circle Law}

\begin{figure}[t!]
\centering
\includegraphics[width=1.0\linewidth]{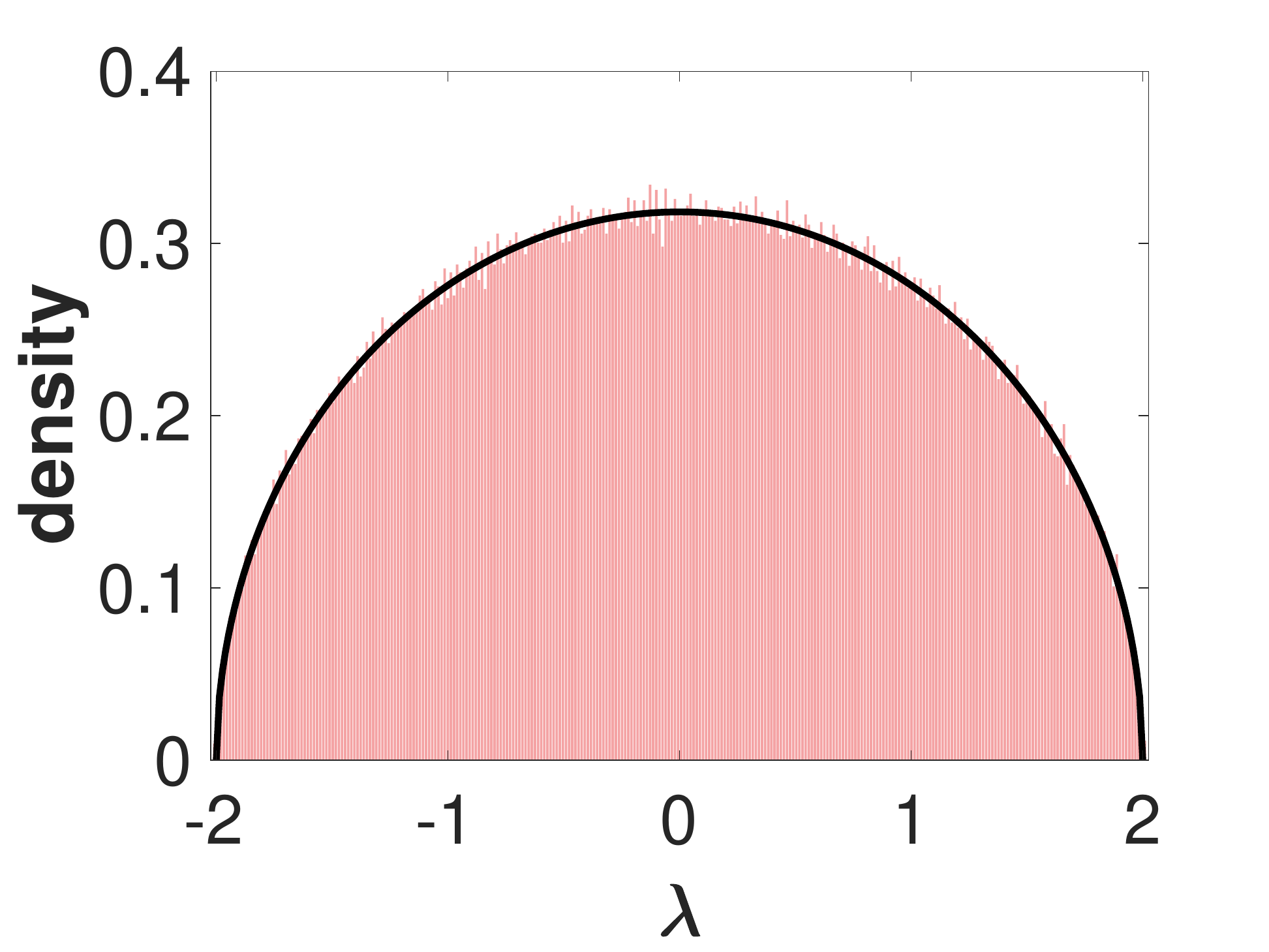} 
\caption{{\bf Wigner semi-circle law for symmetric matrices}. The spectrum of a $N \times N$ symmetric random matrix $A$ whose entries are \emph{i.i.d.} (up to symmetry)
with mean zero and variance $\sigma^2/N$ for $\sigma^2=1$ and $N=1000$ with 100 independent realizations [see Eq.~\eqref{Eq:prob-wigner}]. The solid black line shows analytic predictions of the 
Wigner semi-circle law which is exact in the thermodynamic limit $N \rightarrow \infty$.
 }\label{fig:wigner}
\end{figure}

We start by using the Perturbative Resolvent Method (PRM) to derive the spectrum for the ensemble of real, symmetric random matrices $A$.
We assume that the entries of $A$, denoted $A_{ij}$ ($i,j=1,\ldots, N$), are independent (up to symmetry) with mean and variance of the form
\begin{align}
\expval{A_{ij}} = 0\qc \expval{A_{ij}A_{kl}} = \frac{\sigma^2}{N}(\delta_{ik}\delta_{jl} + \delta_{il}\delta_{jk}).\label{Eq:prob-wigner}
\end{align}
Wigner showed that the spectral density of this ensemble is described by the semi-circle law shown in Fig.~\ref{fig:wigner}~\cite{wigner1993characteristic}. More explicitly, in thermodynamic
limit the spectral density is just 
\begin{align}
\rho_A(x) &= \left\{\begin{array}{cl}
\frac{1}{2\pi\sigma^2}\sqrt{4\sigma^2 -x^2} & \qif \abs{x} \leq 2\sigma\\
0 & \qif \abs{x} > 2\sigma.
\end{array}\right.
\end{align}

\subsection{Resolvent Equations}\label{sec:wigner_reseq}

To calculate this spectral density using the PRM, 
we first construct a system of equations that that will allow us to solve for the Green's function $G_A(z)$ of Eq.~\eqref{Eq:ResolventDef}.
We propose the following system of $N$ equations for $N$ unknown variables $u_i$:
\begin{align}
z u_i &= \sum_{j=1}^NA_{ij}u_j + a_i.\label{Eq:Wigner_LinSys}
\end{align}
We have also introduced the constant $z$ along with $N$ constant auxiliary variables $a_i$.

To see how this system of equation encodes $G_A(z)$,
we ask how the solutions change if we add a small,  independent perturbation $\delta a_i$ to each of the $a_i$.
In the PRM, the response to
these perturbations is characterized by the $N \times N$ ``susceptibility matrix''
\begin{align}
\nu_{ij} &= \pdv{u_i}{a_j}.\label{Eq:susceptibilityWigner}
\end{align}
It is straightforward to show that this susceptibility matrix is equivalent to $G_A(z)$.

To see this, we take the derivative of Eq.~\eqref{Eq:Wigner_LinSys} with respect to $a_k$ to get
\begin{align}
z \nu_{ik} &= \sum_{j=1}^N A_{ij}\nu_{jk} + \delta_{ik}
\end{align}
where $\delta_{ik}$ is just the Kronecker-Delta function.
In matrix form, this set of equations can be written as
\begin{align}
(zI-A)\nu(z) &= I
\end{align}
where $I$ is the $N\times N$ identity matrix.
Equivalently, we can write
\begin{align}
\nu(z) = \frac{1}{zI-A}
\end{align}
which we see is identical to Eq.~\eqref{Eq:ResolventDef},
showing that the Green's function $G_A(z)$ is exactly the susceptibility of the linear equations above.
In particular, we are interested in the trace of the susceptibility
\begin{align}
\bar{\nu} &= \frac{1}{N}\sum_{j=1}^N \nu_{jj}(z)\label{Eq:Wigner_nu}
\end{align}
which allows us to calculate the spectral density using  Eq.~\eqref{Eq:ResolventTrick}.

\subsection{Cavity Expansion}

To calculate the susceptibility, we assume replica symmetry and make use of the zero-temperature cavity method. 
In the cavity method, one relates a system of $N$ equations of the $N$ variables $u_i$ to a system with $N+1$ equations and $N+1$ variables.
By convention, we denote the additional variable by $u_0$ and the additional rows and columns of the matrix $A$ by $A_{0i}$ and $A_{0j}$, respectively. 
In the presence of these new variables,  Eq.~\eqref{Eq:Wigner_LinSys} becomes
\begin{align}
z u_i &= \sum_j A_{ij}u_j + a_i +  A_{i0}u_0 \label{Eq:Wigner_LinSys2}
\end{align}
(Note: indices are implied to range from $1$ to $N$ and terms with $0$-valued indices always be specified explicitly).

From Eq.~\eqref{Eq:prob-wigner}, we know that the extra terms $A_{i0}u_0$ scale as $1/N$.
Therefore, we treat these extra terms as small perturbations to the original equaitons via the auxiliary variables $a_i$,
\begin{align}
\delta a_i = A_{i0}u_0,
\end{align}
and treat them using perturbation theory. 
We also know that the solutions $u_i$ to Eq.~\eqref{Eq:Wigner_LinSys2}  must be related to the solutions $u_{i\setminus 0}$ of Eq.~\eqref{Eq:Wigner_LinSys} (i.e., without the new variable $u_0$) by the perturbative relation 
\begin{align}
\begin{split}
u_i &\approx u_{i\setminus 0} + \sum_j \nu_{ij}\delta a_j\\
& = u_{i\setminus 0} + u_0\sum_j \nu_{ij}A_{j0}
\end{split}
\end{align}
where we have made use of the definition of the susceptibility [Eq.~\eqref{Eq:susceptibilityWigner}].

Now let us turn to additional equation corresponding to the new row in the matrix:
\begin{align}
z u_0 = \sum_j A_{0j}u_j + a_0 + A_{00}u_0
\end{align}
Substituting the perturbative expansion above, we get
\begin{align}
z u_0 = \sum_j A_{0j}u_{j\setminus 0} + u_0\sum_{jk}\nu_{ij}A_{j0}A_{k0}+ a_0 + A_{00}u_0\label{Eq:Wigner_LinSys_Expanded}
\end{align}

\subsection{Approximation via Central Limit Theorem}\label{sec:wigner_CLT}

In the next step in the PRM, we approximate the sum in Eq.~\eqref{Eq:Wigner_LinSys_Expanded} that includes the susceptibility matrix.
In this sum, the elements of the susceptibility matrix are statistically independent of the the new elements of $A$ (those with at least one $0$-valued index).
In addition, this sum includes a very large (order $N$) number of statistically independent terms.
Due to these two properties, it is straightforward to show using the central limit theorem that this sum will be dominated by its mean with respect to the new elements of $A$.
Performing this average, we find that the sum takes the form
\begin{align}
\begin{split}
\sum_{jk}\nu_{ij}A_{j0}A_{k0} &\approx \sum_{jk}\nu_{ij}\expval{A_{j0}A_{k0}}\\
&= \frac{\sigma^2}{N}\sum_{jk}\nu_{ij} \delta_{ij}\\
&= \sigma^2\bar{\nu}
\end{split}
\end{align}
where $\bar{\nu}$ is the trace of the susceptibliity defined in Eq.~\eqref{Eq:Wigner_nu}.

\subsection{Self-Consistency Equation for Susceptibility}

Applying the approximation from the previous section we rewrite Eq.~\eqref{Eq:Wigner_LinSys_Expanded} as
\begin{align}
z u_0 \approx \sum_j A_{0j}u_{j\setminus 0} + u_0\sigma^2\bar{\nu}+ a_0
\end{align}
where we have also dropped the term $A_{00}u_0$ because it is small (order $1/N$).
Rearranging we find the following expression for $u_0$:
\begin{align}
u_0 &= \frac{\sum_j A_{0j}u_{j\setminus 0} + a_0}{1-\sigma^2\bar{\nu}}.
\end{align}
Thus, $u_0$ is a Gaussian random variable. By definition, we know that
\begin{align}
\expval{\nu_{00}} &= \expval{\pdv{u_0}{a_0}} = \frac{1}{1-\sigma^2\bar{\nu}}.
\end{align}
However, since there is nothing special about $u_0$ (i.e., the system self-averages) ,
it is evident that 
\begin{align}
 \bar{v} = \frac{1}{N}\sum_j \nu_{jj} \approx \expval{\nu_{00}}.
\end{align}
This gives us a self-consistency equation for $\bar{v}$ of the form
\begin{align}
\bar{\nu} &= \frac{1}{z-\sigma^2\bar{\nu}}
\end{align}
or equivalently, the quadratic equation
\begin{align}
\sigma^2\bar{\nu}^2 - z\bar{\nu} + 1 &= 0.
\end{align}

\subsection{Spectral Density via Resolvent}

Using the quadratic formula, we get
\begin{align}
\bar{\nu} &= \frac{z\pm\sqrt{z^2-4\sigma^2}}{2\sigma^2}.
\end{align}
To extract the spectrum, we make use of Eq.~\eqref{Eq:ResolventTrick} to relate the spectral density to the imaginary part of the Green's function. 
After substituting $z=x + i 0^+$, it is easy to convince oneself that the only way to get imaginary numbers here is to have $z=x$ with $\abs{x} \leq 2 \sigma$. This yields the expression
\begin{align}
\rho_A(x) &= \left\{\begin{array}{cl}
\frac{1}{2\pi\sigma^2}\sqrt{4\sigma^2 -x^2} & \qif \abs{x} \leq 2\sigma\\
0 & \qif \abs{x} > 2\sigma
\end{array}\right.
\end{align}
which is simply Wigner's semi-circle law.

 \section{Marchenko-Pastur Distribution}

\begin{figure}[t!]
\centering
\includegraphics[width=1.0\linewidth]{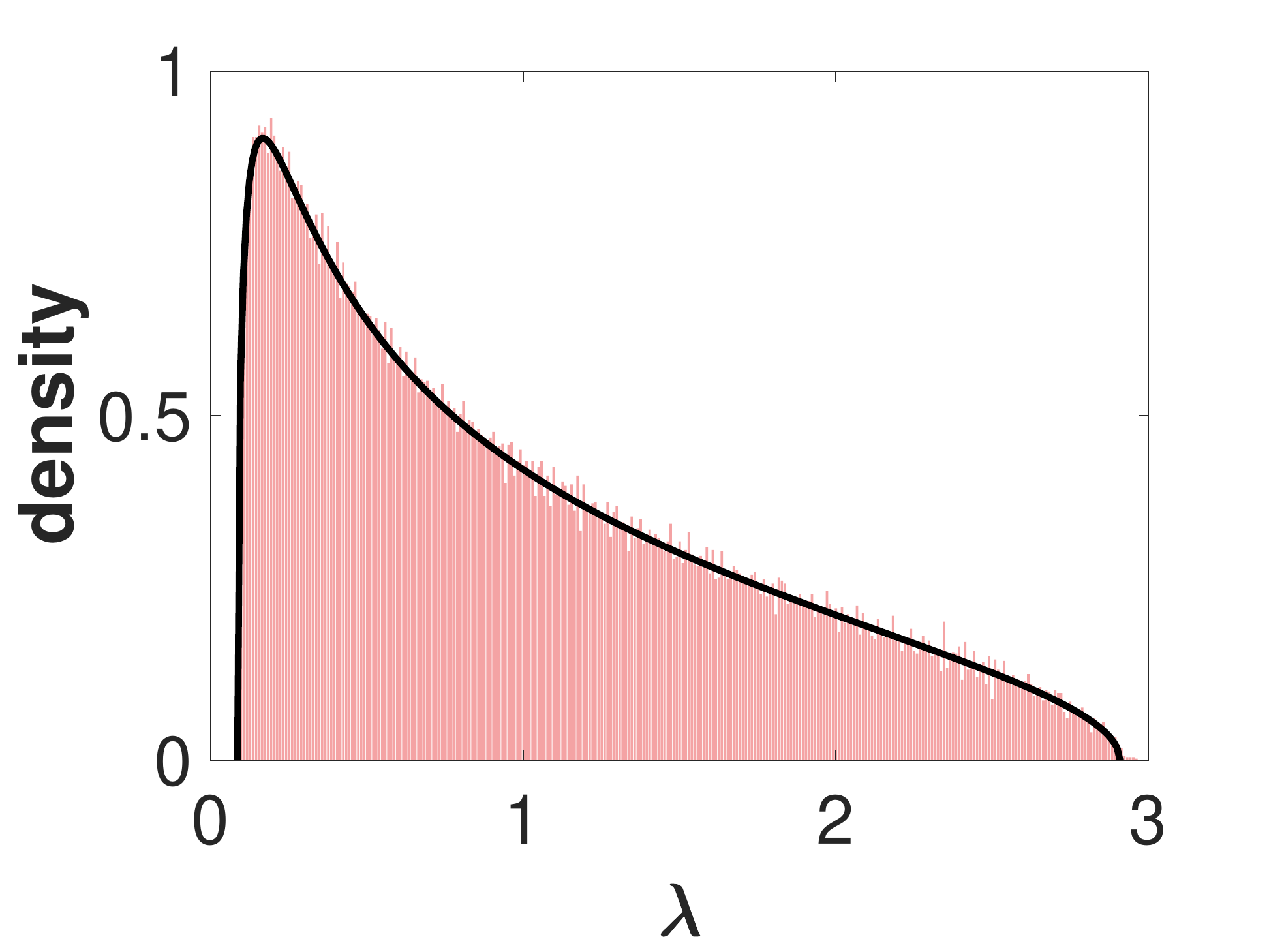} 
\caption{{\bf Marchenko Pastur law for Wishart matrices}. The spectrum of a Wishart random matrix of the form $A=CC^T$ 
where $C$ is an $M \times N$ matrix with entries \textit{i.i.d} entries $C_{i \alpha}$ with mean zero and variance $\sigma^2/N$ for $\sigma=1$, $M=500 $ and $N=1000$ with 100 independent realizations. The solid black line shows analytic predictions of the 
Marchenko-Pastur law [Eq.~\eqref{eq:marchenkopastur}] which is exact in thermodynamic limit $N,M \rightarrow \infty$.
}\label{fig:MP}
\end{figure}

We now use the PRM to derive the Marchenko-Pastur distribution. As before, we define 
a linear system of equations whose susceptibilities are related to the relevant Green's function. We will be interested in the spectral density of the ensemble of
the $M \times M$ Wishart matrices  $A=CC^T$  where $C$ is an $M \times N$ matrix with entries  $C_{i \alpha}$ ($i=1,\ldots, M$ and  $\alpha=1,\ldots,N$) which are drawn from a normal distribution with
mean and variance given by
\begin{align}
\expval{C_{i\alpha}} = 0\qc  \expval{C_{i\alpha}C_{j\beta}} = \frac{\sigma^2}{N}\delta_{ij}\delta_{\alpha\beta} \label{Eq:cstatistics}
\end{align}
Furthermore, we define the ratio
\be
\gamma = {M \over N}.
\ee
We will be interested in the limit $M,N \rightarrow \infty$ with $\gamma$ fixed. A well known result in RMT is that the spectrum of such Wishart matrices is given by the Marchenko-Pastur distribution~\cite{marchenko1967distribution} (see Fig.~\ref{fig:MP}). 
We re-derive this result using the PRM.

\subsection{Resolvent Equations}\label{sec:wishart_reseq}
To apply the PRM, we start with the same system of equations used in the in the previous section to compute the Wigner semi-circle law, Eq.~\eqref{Eq:Wigner_LinSys}.
Inserting the definition of the Wishart matrix $A=CC^T$, 
we get
\begin{align}
zu_i &= \sum_{j\alpha}C_{i\alpha}C_{j\alpha}u_j + a_i\label{eq:wishart_simple}
\end{align}
where $u_i$ are a set of $N$ unknown variables, $z$ is a constant, and $a_i$ are a set of $N$ of constant auxiliary variables (Note: indices represented by Roman letters are implied to range from $1$ to $N$, while indices represented by Greek letters range from $1$ to $M$).
Following the derivation in Sec.~\ref{sec:wigner_reseq}, 
it is clear that the susceptibility matrix
\begin{align}
\nu_{ij}^{(u)} &= \pdv{u_i}{a_j}
\end{align}
is equivalent to the Green's function  $G_A(z)$.

However, to apply the PRM, we require a system of equations that that is linear in the elements of the constituent matrix $C$.
Such a system of equations is found by rewriting Eq.~\eqref{eq:wishart_simple} in terms of an extra set of $M$ unknown variables $v_\alpha$ such that 
\begin{align}
\begin{split}
zu_i &= \sum_\alpha C_{i\alpha}v_\alpha + a_i\\
v_\alpha &= \sum_jC_{j\alpha}u_j + b_\alpha\label{Eq:LinEqWishart}
\end{split}
\end{align}
where we have introduced an extra set of $M$ constant auxiliary variables $b_\alpha$ to complement those in the first equation.
We note that this system of equations is unique as the right hand side of the first equation must depend on the index $i$ and thus the sum must range over the index $\alpha$.
We note that as long as a matrix can be written as a product of matrices such a system of equations can always be uniquely defined (see Sec.~\ref{sec:wishartprod} for an example for a product of four matrices). 

As before, we will be interested in small perturbations around the solutions to these equations.
Therefore, we must consider an expanded set of susceptibilities, 
\begin{align}
\begin{split}
\nu_{ij}^{(u)} &= \pdv{u_i}{a_j}\qc \nu_{\alpha j}^{(v)} = \pdv{v_\alpha}{a_j},\\
\chi_{i\beta}^{(u)} &= \pdv{u_i}{b_\beta}\qc \chi_{\alpha \beta}^{(v)} = \pdv{v_\alpha}{b_\beta},\label{Eq:susWishart}
\end{split}
\end{align}
that measure the extent to which $u_i$ and $v_\alpha$ change in response to a small perturbations in the constants $a_i$ and $b_\alpha$.

Just as $\nu_{ij}^{(u)}$ is the Green's function, the other susceptibility matrices can also be expressed in terms of $z$ and $C$.
Differentiating this system of equations with respect to the auxiliary variables $a_j$ and $b_\alpha$ yields
\begin{align}
\begin{split}
z\nu^{(u)}_{ik} &= \sum_\alpha C_{i\alpha}\nu^{(n)}_{\alpha k} + \delta_{ik}\\
z\chi^{(u)}_{i\beta} &= \sum_\alpha C_{i\alpha}\chi^{(v)}_{\alpha\beta} \\
\nu^{(v)}_{\alpha k} &= \sum_jC_{j\alpha}\nu^{(u)}_{jk} \\
\chi^{(v)}_{\alpha\beta} &= \sum_jC_{j\alpha}\chi^{(u)}_{j\beta} + \delta_{\alpha\beta}
\end{split}
\end{align}
which we can rewrite in matrix form as
\begin{align}
\mqty(zI_N & -C\\
-C^T & I_M)
\mqty(\nu^{(u)} & \chi^{(u)}\\
\nu^{(v)} & \chi^{(v)}) &= 
\mqty(I_N & 0 \\0 & I_M)
\end{align}
where $I_N$ is the $N\times N$ identity matrix and $I_M$ is the $M\times M$ identity matrix.
Inverting this equation yields 
\begin{align}
\mqty(\nu^{(u)} & \chi^{(u)}\\
\nu^{(v)} & \chi^{(v)}) &= \mqty(zI_N & -C\\
-C^T & I_M)^{-1}
\end{align}
We then use standard formulas for inverting block matrices~\cite{lu2002inverses} to get
\begin{align}
\begin{split}
\nu^{(u)} &=  (zI_N- CC^T)^{-1} = \frac{1}{z-A}  \\
\chi^{(u)} &=   (zI_N- CC^T)^{-1}  C  \\
\nu^{(v)} &= C^T  (zI_N- CC^T)^{-1} \\
\chi^{(v)} &= I_M+ C^T (zI_N- CC^T)^{-1}C
\end{split}
\end{align}
Comparing with Eq.~\eqref{Eq:ResolventDef} we see that $\nu^{(u)}$ is exactly the resolvent of the random matrix $A$ as expected.

\subsection{Cavity Expansion}

We once again calculate susceptibilities using the zero temperature cavity method. To do so, we introduce two new variables
$u_0$ and $v_0$, and a new row and column to the matrix $C$ denoted by $C_{0 \alpha}$ and $C_{i 0}$, respectively.  In the presence of these new variables, the original $M+N$ equations are modified to
\begin{align}
\begin{split}
zu_i &= \sum_\alpha C_{i\alpha}v_\alpha + a_i + C_{i0}v_0\\
v_\alpha &= \sum_jC_{j\alpha}u_j + b_\alpha + C_{0\alpha}u_0\label{Eq:LinEqWishart2}
\end{split}
\end{align}

Next, we interpret the terms $C_{i0}v_0$ and $C_{0\alpha}u_0$ as small perturbations to the constants $a_i$ and $b_\alpha$ since the matrix elements of $C$ scale
as $1/N$,
\begin{align}
\delta a_i = C_{i0}v_0 \qc \delta v_\alpha = C_{0\alpha}u_0.
\end{align}
We then perturbatively  relate the solution  to Eq.~\eqref{Eq:LinEqWishart2} in the presence of the new variables, $u_0$ and $v_0$, to the solutions of
Eq.~\eqref{Eq:LinEqWishart} without the new variables, denoted $u_{i \setminus 0}$ and $v_{\alpha \setminus 0}$, using the definitions of the susceptibilities [Eq.~\eqref{Eq:susWishart}],
\begin{align}
\begin{split}
u_i &\approx u_{i\setminus 0} + \sum_j \nu^{(u)}_{ij}C_{j0}v_0 + \sum_\beta \chi^{(u)}_{i\beta} C_{0\beta}u_0\\
v_\alpha &\approx v_{\alpha \setminus 0} + \sum_j \nu^{(v)}_{\alpha j}C_{j0}v_0 + \sum_\beta \chi^{(v)}_{\alpha\beta} C_{0\beta}u_0.
\end{split}
\end{align}

Now we consider the new equations for the two new variables,
\begin{align}
\begin{split}
zu_0 &= \sum_\alpha C_{0\alpha}v_\alpha + a_0 + C_{00}v_0\\
v_0 &= \sum_jC_{j0}u_j + b_0 + C_{00}u_0.\label{Eq:WishartnewEq}
\end{split}
\end{align}
Substituting the expansions into Eq. \ref{Eq:WishartnewEq} gives
\begin{widetext}
\begin{align}
\begin{split}
zu_0 &= \sum_\alpha C_{0\alpha}v_{\alpha \setminus 0} + v_0\sum_{\alpha j}\nu^{(v)}_{\alpha j} C_{0\alpha}C_{j0} + u_0\sum_{\alpha\beta}\chi^{(v)}_{\alpha\beta} C_{0\alpha}C_{0\beta} + a_0 + C_{00}v_0\\
v_0 &= \sum_jC_{j0}u_{j\setminus 0} + v_0\sum_{jk}\nu^{(u)}_{jk}C_{j0}C_{k0}+ u_0\sum_{j\beta}\chi^{(u)}_{j\beta} C_{j0}C_{0\beta} + b_0 + C_{00}u_0.\label{Eq:wishart_expand}
\end{split}
\end{align}
\end{widetext}

\subsection{Approximation via Central Limit Theorem}
Similar to the approximation made in the derivation of the Wigner semi-circle law in Sec.~\ref{sec:wigner_CLT}, we approximate each of the large sums involving the susceptibilities matrices using the Central Limit Theorem.
In particular, it is straightforward to show that the variance of each of these sums is small in the thermodynamic limit $N, M\rightarrow \infty$,
allowing us to approximate each with just its mean with respect to the new row an column  of $C$ (those elements with at least one $0$-valued index).
First, we approximate the the sums involving the square susceptibility matrices $\nu^{(u)}_{jk}$ and $\chi^{(v)}_{\alpha\beta}$ as
\begin{align}
\begin{split}
\sum_{jk}\nu^{(u)}_{jk}C_{j0}C_{k0} &\approx \sum_{jk}\nu^{(u)}_{jk}\expval{C_{j0}C_{k0}} = \sigma^2\bar{\nu}\\
\sum_{\alpha\beta}\chi^{(v)}_{\alpha\beta} C_{0\alpha}C_{0\beta} &\approx \sum_{\alpha\beta}\chi^{(v)}_{\alpha\beta} \expval{C_{0\alpha}C_{0\beta}} = \sigma^2\gamma \bar{\chi}\label{eq:wish_clt}
\end{split}
\end{align}
where we have defined the traces to the susceptibilities as
\begin{align}
\begin{split}
\bar{\nu} &= \frac{1}{N}\sum_j \nu^{(u)}_{jj}\\
\bar{\chi} &= \frac{1}{M}\sum_\alpha \chi^{(v)}_{\alpha\alpha}.
\end{split}
\end{align}
Each term in the other two sums contains a pair of independent elements of $C$, resulting in zero mean,
\begin{align}
\begin{split}
\sum_{\alpha j}\nu^{(v)}_{\alpha j} C_{0\alpha}C_{j0} &\approx \sum_{\alpha j}\nu^{(v)}_{\alpha j} \expval{C_{0\alpha}C_{j0}} = 0\\
\sum_{j\beta}\chi^{(u)}_{j\beta} C_{j0}C_{0\beta} &\approx \sum_{j\beta}\chi^{(u)}_{j\beta} \expval{C_{j0}C_{0\beta}} = 0.\label{eq:wish_cltzero}
\end{split}
\end{align}
As a  result only the two square susceptibility matrices are required for this calculation.

\subsection{Self-Consistency Equations for Susceptibilities}
Applying the approximations from the previous sections to Eq.~\eqref{Eq:wishart_expand} (and dropping the terms $C_{00}v_0$ and $C_{00}u_0$ since there scale as $1/N$) we get
\begin{align}
\begin{split}
zu_0 &= \sum_\alpha C_{0\alpha}v_{\alpha \setminus 0}  + u_0\sigma^2\gamma \bar{\chi} + a_0 \\
v_0 &= \sum_jC_{j0}u_{j\setminus 0} + v_0\sigma^2\bar{\nu} + b_0
\end{split}
\end{align}
which can be rearranged to yield
\begin{align}
\begin{split}
u_0 &= \frac{\sum_\alpha C_{0\alpha}v_{\alpha \setminus 0}+ a_0 }{z-\sigma^2\gamma \bar{\chi}}\\
v_0 &= \frac{\sum_jC_{j0}u_{j\setminus 0} + b_0}{1-\sigma^2\bar{\nu}}.
\end{split}
\end{align}
From the cavity construction, we know that the traces of the susceptibility self-averaging, allowing us to write
\begin{align}
\begin{split}
\bar{\nu} &\approx \expval{\nu^{(u)}_{00}} = \expval{\pdv{u_0}{a_0}} = \frac{1}{z-\sigma^2\gamma \bar{\chi}}\\
\bar{\chi} &\approx \expval{\chi^{(v)}_{00}} = \expval{\pdv{v_0}{b_0}} = \frac{1}{1-\sigma^2\bar{\nu}}.
\end{split}
\end{align}
We combine these equations to get 
\begin{align}
\bar{\nu} &= \frac{1}{z-\sigma^2\frac{1}{1-\gamma\sigma^2\bar{\nu}} }
\end{align}
which gives us a quadratic equation for $\bar{\nu}$,
\begin{align}
z\gamma \sigma^2 \bar{\nu}^2 - \qty[z + \sigma^2( \gamma-1)]\bar{\nu} +1 =0.
\end{align}

\subsection{Spectral Density via Resolvent}

Solving the quadratic equation, we obtain
\begin{align}
\bar{\nu}(z) =  \frac{z + \sigma^2 (\gamma-1)}{2 z\gamma \sigma^2} \pm \frac{\sqrt {\qty[z + \sigma^2( \gamma-1)]^2 -4 z\gamma \sigma^2}}{2 z\gamma \sigma^2}.
\label{Eq:MPnu}
\end{align}

To extract the spectral density, we make use of Eq.~\eqref{Eq:ResolventTrick}. Notice that the imaginary part of the first term in the sum above for $z=x+i0^+$
is simply
\be
\lim_{\epsilon \rightarrow 0^+} {1 \over 2 \pi}(1-\gamma^{-1}) {\epsilon \over x^2 +\epsilon^2},
\ee
which we recognize as the definition of delta function at zero. For this reason, this term accounts for the spectral weight of the zero eigenvalues and hence is non-zero only
if the Wishart matrix is not full rank (i.e. $\gamma >1$).

Now consider the second term in Eq.~\eqref{Eq:MPnu}. In order to have a non-zero imaginary part of $\bar{\nu}$ when $z=x+ i0^+$, the discriminant must be negative. 
This allows us to calculate the upper and lower bounds for where $\rho(x)$ is non-zero by setting the discriminant in Eq.~\eqref{Eq:MPnu} to zero with $z=x$,
\begin{align}
\qty[x + \sigma^2( \gamma-1)]^2 -4 x\gamma \sigma^2 = 0,
\end{align}
yielding,
\begin{align}
\begin{split}
x_{\min} &=  (\gamma + 1)\sigma^2 - 2 \sqrt{\gamma}\sigma^2 \\
x_{\max} &=  (\gamma + 1)\sigma^2 + 2 \sqrt{\gamma}\sigma^2.
\end{split}
\end{align}

\begin{widetext}
Combining this with Eq.\eqref{Eq:MPnu}, the contribution of this term to the spectral density is non-zero only on a finite region $[x_{\min}, x_{\max}]$.  Together,
these observations yield the following expression for the spectral density
\begin{align}
\rho_A(x) &= \left\{\begin{array}{cl}
(1-\gamma^{-1})\delta(x)+  \frac{1}{2 \pi x \gamma \sigma^2}\sqrt{(x-x_{\min})(x_{\max}-x)}, &\qif \gamma>1 \\
 \frac{1}{2 \pi x \gamma \sigma^2}\sqrt{(x-x_{\min})(x_{\max}-x)},  & \qif \gamma \le 1 .
\end{array}\right.\label{eq:marchenkopastur}
\end{align}
This is exactly the Marchenko-Pastur distribution.
\end{widetext}
\section{Wishart Product Matrices}\label{sec:wishartprod}

Next, we use the PRM to derive the spectrum for a special case of Wishart product matrices \cite{burda2010eigenvalues, dupic2014spectral}.
In the general case, these matrices take the form $A=CC^T$ where $C=B_1B_2\times\cdots \times B_n$ where each matrix $B_i$ can be a rectangular matrix of a different size (such that the matrix multiplications are valid).
In this section, we consider the the spectral density of the ensemble of Wishart product matrices $A=CBB^TC^T$ where $C$ and $B$ are both $N\times N$ matrices whose entries, $B_{ij}$ and $C_{ij}$, respectively, are independently drawn from normal distributions with mean and variances given by
\begin{gather}
\begin{gathered}
\expval{B_{ij}} = 0\qc \expval{B_{ij}B_{kl}} = \frac{\sigma_B^2}{N}\delta_{ik}\delta_{jl}\\
\expval{C_{ij}} = 0\qc \expval{C_{ij}C_{kl}} = \frac{\sigma_C^2}{N}\delta_{ik}\delta_{jl}\\
\expval{B_{ij}C_{kl}} = 0.
\end{gathered}
\end{gather}

\subsection{Resolvent Equations}
Following Sec.~\ref{sec:wigner_reseq}, we begin by inserting the definition for $A$ into Eq.~\eqref{Eq:Wigner_LinSys},
\begin{align}
z u_i &= \sum_{jklm}C_{ij}B_{jk}B_{lk}C_{ml}u_m + a_i\label{eq:wishprod_oneeq}
\end{align}
where $u_i$ $(i=1,\cdots, N)$ are a set of unknown variables, $z$ is a constant, and $a_i$ are a set of constant auxiliary variables.
Just as the previous section, it is clear that the Green's function $G_A(z)$ is given by the susceptibility matrix
\begin{align}
\nu^{(u)}_{ij} = \pdv{u_i}{a_j}.
\end{align}

Next, we follow the setup described for product matrices in Sec.~\ref{sec:wishart_reseq}, defining a new set of unknown variables,  $v_i$, $w_i$, and $x_i$ $(i=1,\cdots, N)$, allowing us to decompose Eq.~\eqref{eq:wishprod_oneeq} such that each equation is linear in the elements of the random matrices $C$ and $B$,
\begin{align}
\begin{split}
z u_i &= \sum_jC_{ij}v_j + a_i\\
v_i &= \sum_jB_{ij}w_j + b_i\\
w_i &= \sum_jB_{ji}x_j + c_i\\
x_i &= \sum_jC_{il}u_i + d_i.\label{eq:wishprod_linsys}
\end{split}
\end{align}
In each equation of the additional three equations, we have also introduced a new set of constant auxiliary variables, $b_i$, $c_i$, and $d_i$. 

In order to explore small perturbations about the solutions to these equations, we define all possible susceptibility matrices with respect to the auxiliary variables,
\begin{align}
\begin{split}
\nu^{(u)}_{ij} &= \pdv{u_i}{a_j}, \nu^{(v)}_{ij} = \pdv{v_i}{a_j}, \nu^{(w)}_{ij} = \pdv{w_i}{a_j}, \nu^{(x)}_{ij} = \pdv{x_i}{a_j}\\
\chi^{(u)}_{ij} &= \pdv{u_i}{b_j} , \chi^{(v)}_{ij} = \pdv{v_i}{b_j} , \chi^{(w)}_{ij} = \pdv{w_i}{b_j} , \chi^{(x)}_{ij} = \pdv{x_i}{b_j}\\
\phi^{(u)}_{ij} &= \pdv{u_i}{c_j} , \phi^{(v)}_{ij} = \pdv{v_i}{c_j} , \phi^{(w)}_{ij} = \pdv{w_i}{c_j} , \phi^{(x)}_{ij} = \pdv{x_i}{c_j}\\
\omega^{(u)}_{ij} &= \pdv{u_i}{d_j} , \omega^{(v)}_{ij} = \pdv{v_i}{d_j} , \omega^{(w)}_{ij} = \pdv{w_i}{d_j} , \omega^{(x)}_{ij} = \pdv{x_i}{d_j}.
\end{split}
\end{align}
As in the previous section, only a small subset of these susceptibilities will be relevant.
%
%

\subsection{Cavity Expansion}

Next, we begin the zero-cavity method by introducing a four variables, $u_0$, $v_0$, $w_0$, and $x_0$, and a new row and column to each matrices denoted by $C_{i0}$, $C_{0j}$, $B_{i0}$, and $B_{0j}$. In the presence of these new variables, the original $4N$ equations [Eq.~\eqref{eq:wishprod_linsys}] become
\begin{align}
\begin{split}
z u_i &= \sum_jC_{ij}v_j + a_i + C_{i0}v_0\\
v_i &= \sum_jB_{ij}w_j + b_i  + B_{i0}w_0\\
w_i &= \sum_jB_{ji}x_j + c_i + B_{0i}x_0\\
x_i &= \sum_jC_{ji}u_j + d_i + C_{0i}u_0.
\end{split}
\end{align}

From here, we interpret the extra terms as small (order $1/N$) perturbations to the auxiliary variables,
\begin{align}
\begin{split}
\delta a_i = C_{i0}v_0\qc \delta b_i = B_{i0}w_0\\
\delta c_i = B_{0i}x_0 \qc \delta d_i = C_{0i}u_0,
\end{split}
\end{align}
allowing us to perturbatively relate the solutions of these equations with the new variables to those of Eq.~\eqref{eq:wishprod_linsys}, denoted $u_{i\setminus 0}$, $v_{i\setminus 0}$, $w_{i\setminus 0}$, $x_{i\setminus 0}$. This gives us
\begin{widetext}
\begin{align}
\begin{split}
u_i &\approx u_{i\setminus 0} + \sum_j\nu^{(u)}_{ij}C_{i0}v_0 + \sum_j\chi^{(u)}_{ij}B_{j0}w_0 + \sum_j \phi^{(u)}_{ij}B_{0j}x_0 + \sum_j \omega^{(u)}_{ij}C_{0j}u_0\\
v_i &\approx v_{i\setminus 0} + \sum_j\nu^{(v)}_{ij}C_{i0}v_0 + \sum_j\chi^{(v)}_{ij}B_{j0}w_0 + \sum_j \phi^{(v)}_{ij}B_{0j}x_0 + \sum_j \omega^{(v)}_{ij}C_{0j}u_0\\
w_i &\approx w_{i\setminus 0} + \sum_j\nu^{(w)}_{ij}C_{i0}v_0 + \sum_j\chi^{(w)}_{ij}B_{j0}w_0 + \sum_j \phi^{(w)}_{ij}B_{0j}x_0 + \sum_j \omega^{(w)}_{ij}C_{0j}u_0\\
x_i &\approx x_{i\setminus 0} + \sum_j\nu^{(x)}_{ij}C_{i0}v_0 + \sum_j\chi^{(x)}_{ij}B_{j0}w_0 + \sum_j \phi^{(x)}_{ij}B_{0j}x_0 + \sum_j \omega^{(x)}_{ij}C_{0j}u_0.\label{eq:wishprod_exp}
\end{split}
\end{align}
\end{widetext}

Next, we will consider the additional equations for the the new variables, given by
\begin{align}
\begin{split}
z u_0 &= \sum_jC_{0j}v_j + a_0+ C_{00}v_0\\
v_0 &= \sum_jB_{0j}w_j + b_0  + B_{00}w_0\\
w_0 &= \sum_jB_{j0}x_j + c_0 + B_{00}x_0\\
x_0 &= \sum_jC_{j0}u_j + d_0 + C_{00}u_0.\label{eq:wishprod_zero}
\end{split}
\end{align}

\subsection{Approximation via Central Limit Theorem}

Following the procedure described in the previous sections, 
we substitute the expansions in Eq.~\eqref{eq:wishprod_exp} into the equations for the new variables in Eq.~\eqref{eq:wishprod_zero}.
The resulting set of equations contains many sums over large numbers of random variables.
In particular, for each sum containing one of the susceptibility matrices, we make use of the fact that the susceptibilities are statistically independent of the new rows and columns of $C$ and $B$.
Using the central limit theorem, each sum can be shown to be dominated by its mean.
Four of these sums can be shown to be nonzero, following the same form as those in Eq.~\eqref{eq:wish_clt},
\begin{align}
\sum_{jk}\nu^{(u)}_{jk}C_{j0}C_{k0} &\approx \sigma_C^2\bar{\nu}\\
\sum_{jk}\omega^{(v)}C_{0j}C_{0k} &\approx \sigma_C^2 \bar{\omega}\\
\sum_{jk}\phi^{(w)}B_{0j}B_{0k} &\approx \sigma_B^2 \bar{\phi}\\
\sum_{jk}\chi^{(x)}_{ij}B_{j0}B_{k0} &\approx \sigma_B^2 \bar{\chi}
\end{align}
where we have defined the traces of the susceptibilities matrices as
\begin{gather}
\begin{split}
\bar{\nu} = \frac{1}{N}\sum_j \nu^{(u)}_{jj}\qc \bar{\chi} = \frac{1}{N}\sum_j \chi^{(x)}_{jj}\\
\bar{\phi} = \frac{1}{N}\sum_j \phi^{(w)}_{jj} \qc \bar{\omega} = \frac{1}{N}\sum_j \omega^{(v)}_{jj}.
\end{split}
\end{gather}
It can easily be shown that the remaining twelve sums containing one of the susceptibilities are all similar in form to those in Eq.~\eqref{eq:wish_cltzero} and are therefore approximately zero in the thermodynamic limit $N\rightarrow \infty$.

\subsection{Self-Consistency Equations for Susceptibilities}

Applying the approximations from the previous section and neglecting the additional terms of order $1/N$, Eq.~\eqref{eq:wishprod_zero} becomes
\begin{align}
\begin{split}
z u_0 &\approx \sum_jC_{0j}v_{j\setminus 0}  + u_0\sigma_C^2\bar{\omega} + a_0 \\
v_0 &\approx \sum_jB_{0j}w_{j\setminus 0} + x_0\sigma_B^2\bar{\phi} + b_0\\
w_0 &\approx \sum_jB_{j0}x_{j\setminus 0} + w_0\sigma_B^2\bar{\chi} + c_0 \\
x_0 &\approx \sum_jC_{j0}u_{j\setminus 0} + v_0\sigma_C^2\bar{\nu} + d_0.
\end{split}
\end{align}
Solving these equations for the new variables, we find
\begin{widetext}
\begin{align}
\begin{split}
u_0 &= \frac{\sum_jC_{0j}v_{j\setminus 0}+a_0}{z-\sigma_C^2\bar{\omega}}\\
v_0 &= \frac{\sum_jB_{0j}w_{j\setminus 0} + b_0 + \sigma_B^2\bar{\phi}\qty(\sum_jC_{j0}u_{j\setminus 0} + d_0)}{1-\sigma_B^2\sigma_C^2\bar{\nu}\bar{\phi}}\\
w_0 &= \frac{\sum_jB_{j0}x_{j\setminus 0}  + c_0}{1-\sigma_B^2\bar{\chi}}\\
x_0 &=  \frac{\sigma_C^2\bar{\nu}\qty(\sum_jB_{0j}w_{j\setminus 0} + b_0) + \sum_jC_{j0}u_{j\setminus 0} + d_0}{1-\sigma_B^2\sigma_C^2\bar{\nu}\bar{\phi}}.
\end{split}
\end{align}
\end{widetext}
Next, we approximate the traces of the susceptibility matrices as the average over a single element, giving us a set of self-consistent equations
\begin{align}
\begin{split}
\bar{\bar{\nu}} &\approx \expval{\nu^{(u)}_{00}} = \expval{\pdv{u_0}{a_0}} =  \frac{1}{z-\sigma_C^2\bar{\omega}}\\
\bar{\chi} &\approx \expval{\chi^{(x)}_{00}} = \expval{\pdv{x_0}{b_0}} = \frac{\sigma_C^2\bar{\nu}}{1-\sigma_B^2\sigma_C^2\bar{\nu}\bar{\phi}} \\
\bar{\phi} &\approx \expval{\phi^{(w)}_{00}} = \expval{\pdv{w_0}{c_0}} = \frac{1}{1-\sigma_B^2\bar{\chi}}\\
\bar{\omega} &\approx \expval{\bar{\omega}^{(v)}_{00}} = \expval{\pdv{v_0}{d_0}} =\frac{\sigma_B^2\bar{\phi}}{1-\sigma_B^2\sigma_C^2\bar{\nu}\bar{\phi}}.
\end{split}
\end{align}
Combining these equations, we arrive at a cubic equation for $\bar{\nu}$ of the form
\begin{align}
\sigma_B^2\sigma_C^2z^2\bar{\nu}^3 - z\bar{\nu} + 1 &= 0.
\end{align}

\subsection{Spectral Density via Resolvent}
To solve this cubic equation, we first rewrite it in terms of $z\bar{\nu}$,
\begin{align}
0 &= (z\bar{\nu})^3 - \frac{z}{\sigma_B^2\sigma_C^2}(z\bar{\nu}) + \frac{z}{\sigma_B^2\sigma_C^2},
\end{align}
allowing to express the three solutions in general form as 
\begin{align}
\begin{split}
z\bar{\nu}^{(1)} &= S+T\\
z\bar{\nu}^{(2)} &= -\frac{1}{2}(S+T) + \frac{1}{2}i\sqrt{3}(S-T)\\
z\bar{\nu}^{(2)} &= -\frac{1}{2}(S+T) - \frac{1}{2}i\sqrt{3}(S-T)
\end{split}
\end{align}
where
\begin{align}
\begin{split}
S &= \sqrt[3]{R+\sqrt{D}}\\
T &= \sqrt[3]{R-\sqrt{D}}\\
D &= R^2 - Q^3
\end{split}
\end{align}
with
\begin{align}
\begin{split}
Q &= \frac{1}{3}\frac{z}{\sigma_B^2\sigma_C^2}\\
R &= -\frac{1}{2}\frac{z}{\sigma_B^2\sigma_C^2}.
\end{split}
\end{align}
In writing these expressions, we have made us of standard mathematical identities for the roots of a cubic equation.

We then make use of Eq.~\eqref{Eq:ResolventTrick} to extract the spectral density.
After substituting $z=x+i0^+$, it is possible to convince oneself that spectral density is given by the imaginary part of solution $\bar{\nu}^{(2)}$ when the polynomial discriminant $D$ is positive. 
To solve for the the bounds of the distribution, we solve for the roots of $D$ after setting $z=x$,
\begin{align}
\begin{split}
D &= \frac{1}{4}\qty(\frac{x}{\sigma_B^2\sigma_C^2})^2 - \frac{1}{27}\qty(\frac{x}{\sigma_B^2\sigma_C^2})^3 = 0.
\end{split}
\end{align}
The limiting eigenvalues are then
\begin{align}
\begin{split}
x_{\min} &= 0\\
x_{\max}  &= \frac{27}{4}\sigma_B^2\sigma_C^2.
\end{split}
\end{align}

\begin{figure}[t!]
\centering
\includegraphics[width=1.0\linewidth]{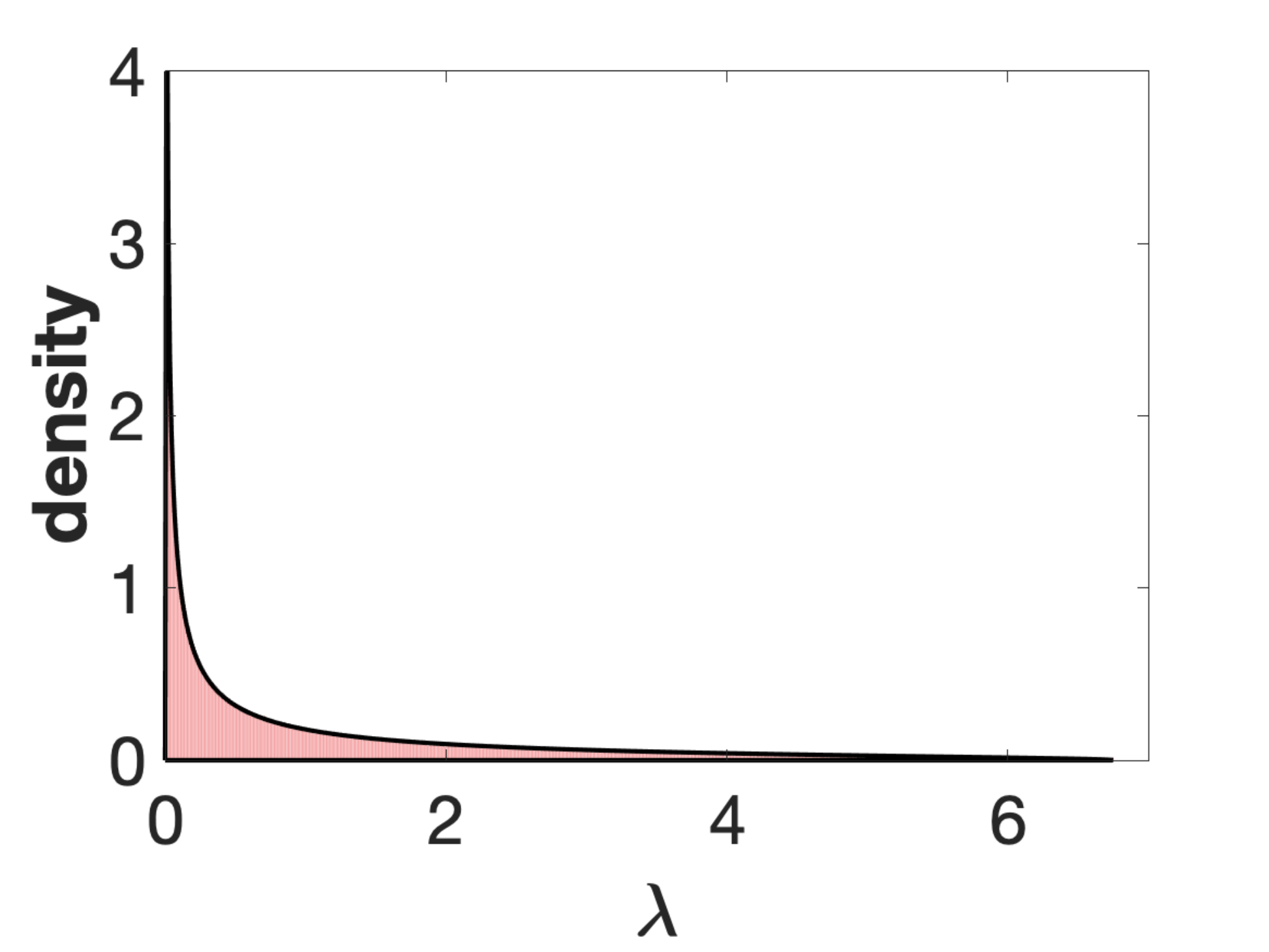} 
\caption{The spectrum of a Wishart product matrix of the form $A=(CB)(CB)^T$ 
where $C$ and $B$  are  $N \times N$ matrices with entries \textit{i.i.d} with mean zero and variances $\sigma^2_C/N$ and $\sigma^2_B/N$, respectively, where $\sigma^2_C=\sigma^2_B=1$ and $N=1000$ with 100 independent realizations. The solid black line shows the analytic prediction of Eq.~\eqref{eq:Wishart_product} which is exact in thermodynamic limit $N \rightarrow \infty$.
}\label{fig:ProductWishart}
\end{figure}

Finally, the spectral density is given by
\begin{align}\label{eq:Wishart_product}
\rho(x) &= \frac{\sqrt{3}}{2\pi x}\qty[S_+\qty(\frac{x}{\sigma_B^2\sigma_C^2}) - S_-\qty(\frac{x}{\sigma_B^2\sigma_C^2})]
\end{align}
where
\begin{align}
S_{\pm}(a) &= \sqrt[3]{\frac{1}{2}a\pm\sqrt{\frac{1}{27}a^2\qty(\frac{27}{4}-a)}}.
\end{align}
Numerical checks of these expression is show in Figure \ref{fig:ProductWishart}.

\section{The Circle and Elliptic Laws}

\begin{figure}[t!]
\centering
\includegraphics[width=1.0\linewidth]{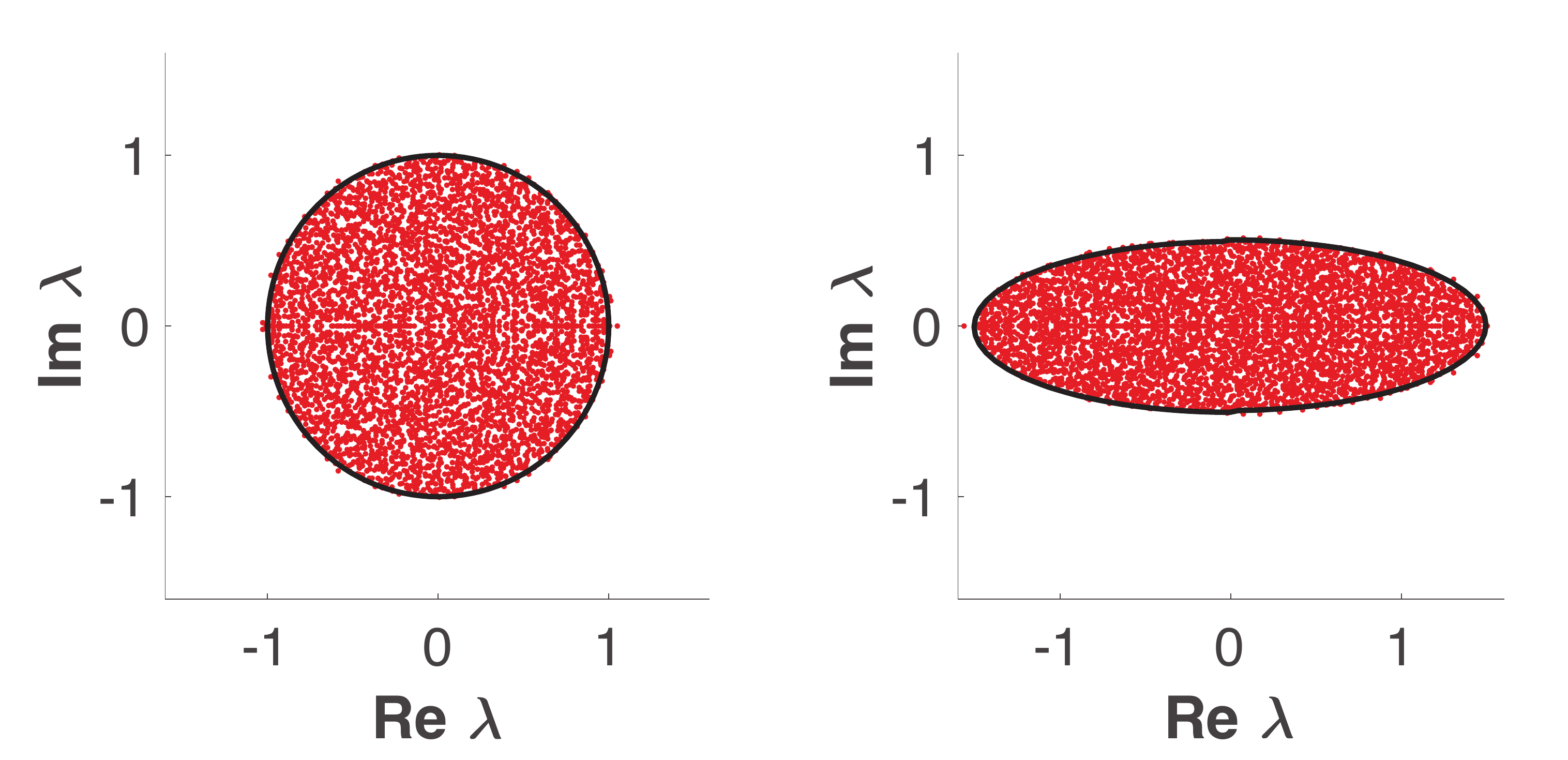} 
\caption{{\bf Circle and elliptic law for real matrices}.The eigenvalues of a real  $N \times N$ matrix $A$ with entries $A_{ij}$ with mean zero
and $\<A_{ij}A_{kl}\>= {\sigma^2 \over N} \delta_{ik}\delta_{jl} + {\zeta \sigma^2 \over N} \delta_{il}\delta_{jk}$. (Left) Spectrum for a matrix with
 $N=1000$, $\sigma =1$, and $\gamma =0$. The solid black line shows analytic predictions of the 
circle law which is exact in thermodynamic limit $N \rightarrow \infty$. (Right) The spectrum for a matrix with $N=1000$, $\sigma =1$, and $\gamma =0.5$.
The solid black line shows analytic predictions of the 
elliptic law which is exact in thermodynamic limit $N \rightarrow \infty$. There are 5 independent realizations in both figures.
}\label{fig:CircleLaw}
\end{figure}

In the previous sections, we considered symmetric matrices which have real eigenvalues. 
We now show how the PRM
can also be used to calculate the spectral density of non-symmetric real matrices where the eigenvalues have both real and imaginary
parts. 
We consider the ensemble of real $N \times N$ matrices $A$ with elements $A_{ij}$ such that 
\begin{align}
\expval{A_{ij}} = 0\qc  \expval{A_{ij}A_{kl}} = \frac{\sigma^2}{N}\delta_{ij}\delta_{kl} + \frac{\zeta\sigma^2}{N}\delta_{il}\delta_{jk}.\label{Eq:Arealstats}
\end{align}
The spectral densities of such ensembles are known to be described by the Girko circle law when $\zeta=0$ \cite{girko1985circular} and more generally,
the elliptic law when $\zeta \neq 0$ (see Fig.~\ref{fig:CircleLaw}).

\subsection{Resolvent Equations}

We now derive both these spectral densities using the PRM. Inspired by Eq.~\eqref{Eq:ZeeH}, we consider the $2N \times 2N$ dimensional matrix
\begin{align}
H(z) &= \mqty(0 & A-zI \\
z^*I-A^* & 0)
\end{align}
where $z$ is a constant and $I$ is the $N\times N$ identity matrix.
In analogy to Eq.~\eqref{Eq:Wigner_LinSys}, we construct a system of $2N$ real equations for the variables $x_i$ and $y_i$ ($i=1, \ldots N$) of the form
\begin{align}
\eta  \mqty(\vec{x} \\
\vec{y})&= H(z)\mqty(\vec{x} \\
\vec{y}) + \mqty(\vec{a} \\
\vec{b})
\end{align}
where $\eta$ is a constant and $a_i$ and $b_i$ are $2N$  real-valued auxiliary variables. 
We rewrite these equations in component form as
\begin{align}
\begin{split}
\eta x_i &= \sum_jA_{ij}y_j -zy_i + a_i\\
\eta y_i &= -\sum_jA^*_{ij}x_j +z^* x_i + b_i.\label{Eq:Systemcircle}
\end{split}
\end{align}
As before, we define a set of susceptibilities, 
\begin{align}
\begin{split}
\chi_{ij}^{(x)} &= \pdv{x_i}{a_j}\qc \nu_{i j}^{(y)} = \pdv{y_i}{a_j},\\
\nu_{ij}^{(x)} &= \pdv{x_i}{b_j}\qc \chi_{ij}^{(y)} = \pdv{y_i}{b_j},\label{Eq:suscircle}
\end{split}
\end{align}
that measure how $x_i$ and $y_j$ change in response to a small changes in the constants $a_i$ and $b_j$.
Taking the derivative of Eq.~\eqref{Eq:Systemcircle}  with respect to $b_k$ and $a_k$ gives
\begin{align}
\begin{split}
\eta \chi_{ik}^{(x)} &= \sum_jA_{ij}\nu_{jk}^{(y)} - z\nu_{ik}^{(y)} + \delta_{ik}\\
\eta \nu_{ik}^{(x)} &= \sum_jA_{ij}\chi_{jk}^{(y)} - z\chi_{ik}^{(y)}\\
\eta \nu_{ik}^{(y)} &= -\sum_jA^*_{ij}\chi_{jk}^{(x)} + z^* \chi_{ik}^{(x)}\\
\eta \chi_{ik}^{(y)} &= -\sum_jA^*_{ij}\nu_{jk}^{(x)} +z^* \nu_{ik}^{(x)} + \delta_{ik}.
\end{split}
\end{align}
We then rewrite this in matrix form as
\begin{align}
\mqty(\eta I & zI-A\\
-z^*I+A^*   & \eta I )
\mqty(\chi^{(x)} & \nu^{(x)}\\
\nu^{(y)} & \chi^{(y)}) &= \mqty(I & 0 \\ 0 & I)
\end{align}

Notice that the equations for the susceptibilities do not depend on the value of $a_i$ and $b_i$. This implies that they are independent of the exact values of $a_i$ and $b_i$
and should be valid even for special case where all the constants are chosen to be identical, $a_i=b_i=c$ for all $i$. 
In this case, there
is a hidden symmetry in our system of linear equations, Eq.~\eqref{Eq:Systemcircle}, implying that the four susceptibilities are not independent, but instead
related by complex conjugation. In light of these observations, it is easy to convince oneself that we must have 
 $\chi^{(x)}=\chi^{(y)}=\chi$, $\nu^{(x)}=\nu$ and $\nu^{(y)}=-\nu^*$ where $\nu^*$ denotes the Hermitian conjugate of $\nu$. 
This implies that we can rewrite our equations in the simpler form
\begin{align}
\mqty(\eta I & z I-A\\
-z^* I +A^*  & \eta I )
\mqty(\chi & \nu\\
-\nu^* & \chi) &= \mqty(I & 0 \\ 0 & I).
\end{align}
Inverting this equation we see that the susceptibility matrices correspond exactly to the complex Green's function of Eq.~\eqref{Eq:GreenComplex},
\begin{align}
\mathcal{G}_A(\eta) &= \mqty(\chi & \nu\\
-\nu^* & \chi).
\end{align}
We can use these expressions to calculate the spectral function of $A$ using Eq.~\eqref{Eq:density-complex}:
\begin{align}
\rho_A(x,y) &=-{1 \over \pi} \partial_{z^*}\eval{\qty[ {1 \over N}\Tr \, \nu^*(z, z^*)]}_{\eta=0} \\
&={1 \over \pi} \partial_{z^*}\eval{\qty[ {1 \over N}\Tr \, \nu^{(y)}(z, z^*)]}_{\eta=0}
\label{Eq:density-complex}
\end{align}

\subsection{Susceptibilities via Cavity Method}

Once again we calculate the susceptibilities using the cavity method. 
To do so, we introduce two additional variables $x_0$ and
$y_0$ and an additional row $A_{0j}$ and column $A_{i0}$ to the matrix $A$. 
With the addition of these new variables, Eq. \ref{Eq:Systemcircle} for $x_i$ and $y_i$ is modified to
\begin{align}
\begin{split}
\eta x_i &= \sum_jA_{ij}y_j -zy_i + a_i + A_{i0} y_0\\
\eta y_i &= -\sum_jA^*_{ij}x_j + z^* x_i + b_i - A_{i0}^* x_0.\label{Eq:Systemcircle2}
\end{split}
\end{align}
As before, we treat the effect of these two extra variables as small perturbations since the additional terms in the
equation above scale as $1/N$. 
In particular, we relate the solutions to Eq.~\eqref{Eq:Systemcircle2} (denoted by $x_i$ and $y_i$) to the solutions to Eq.~\eqref{Eq:Systemcircle2} without the new variables $x_0$ and $y_0$ (denoted by $x_{i\setminus 0}$ and $y_{i\setminus 0}$)
using the definition of the susceptibilities [Eq.~\eqref{Eq:suscircle}],
\begin{align}
\begin{split}
x_j &= x_{j \setminus 0} - \sum_k \nu_{jk}^{(x)}A_{k0}^* x_0 + \sum_k \chi_{jk}^{(x)}y A_{k0}y_0\\
y_j &= y_{j \setminus 0} - \sum_k \chi_{jk}^{(y)}A_{k0}^* x_0 + \sum_k \nu_{jk}^{(y)} A_{k0}y_0,
\end{split}
\end{align}
where in writing this we have made use of the relationships between the various susceptibilities noted above.

In the cavity method, there are also two additional equations for the new variables $x_0$ and $y_0$ given by
\begin{align}
\begin{split}
\eta x_0 &= \sum_jA_{0j}y_j -zy_0 + a_0 + A_{00} y_0\\
\eta y_0 &= -\sum_jA^*_{0j}x_j + z^* x_0 + b_0 - A_{00}^* x_0.
\end{split}
\end{align}
Substituting in the expansions above gives
\begin{widetext}
\begin{align}
\begin{split}
\eta x_0 &= \sum_jA_{0j}y_{j \setminus 0} -\sum_{jk} \chi_{jk}^{(y)}A_{0j}A_{k0}^* x_0 + \sum_{jk} \nu^{(y)}_{jk}A_{0j} A_{k0}y_0 -zy_0 + a_0 + A_{00} y_0\\
\eta y_0 &= -\sum_jA^*_{0j}x_{j \setminus 0} + \sum_{jk} \nu_{jk}^{(x)}A^*_{0j}A_{k0}^* x_0 - \sum_{jk} \chi_{jk}^{(x)}A^*_{0j} A_{k0}y_0 + z^* x_0 + b_0 - A_{00}^* x_0.
\end{split}
\end{align}
\end{widetext}

Next, we approximate each sum containing a susceptibility matrix with its mean (with respect to the new row and column of $A$) using the central limit theorem  in the large $N$ limit (see previous sections). 
Using Eq.~\eqref{Eq:Arealstats} and noting that $A^*_{ij}= A_{ji}$ for real matrices and relationships between susceptibilities, we get
\begin{align}
\begin{split}
\eta x_0 &= \sum_jA_{0j}y_{j \setminus 0} -x_0\sigma^2\bar{\chi}^{(y)} + y_0\zeta\sigma^2\bar{\nu}^{(y)} -zy_0 + a_0 \\
\eta y_0 &= -\sum_jA_{j0}x_{j \setminus 0} +  x_0\zeta\sigma^2\bar{\nu}^{(x)} -  y_0\sigma^2\bar{\chi}^{(x)} + z^* x_0 + b_0,\label{Eq:cavity-full-ellipse}
\end{split}
\end{align}
where the bar on the susceptibility denotes the trace. We are now in a position to solve for $\bar{\nu}^{(y)}$ and calculate
the spectral density.


\subsection{Circle Law}

Let us first focus on the special case where the entries are completely decorrelated with $\zeta=0$. In this case, we know that the spectrum is described by the Girko circle law. 
To derive this, we use Eq.~\eqref{Eq:cavity-full-ellipse} with $\zeta=0$ (noting that $\bar{\chi}^{(x)}=\bar{\chi}^{(y)}=\bar{\chi}$),
\begin{align}
\begin{split}
\eta x_0 &= \sum_jA_{0j}y_{j \setminus 0} - x_0\sigma^2\bar{\chi}  -zy_0 + a_0 \\
\eta y_0 &= -\sum_jA_{j0}x_{j \setminus 0}  -  y_0\sigma^2\bar{\chi}+ z^* x_0 + b_0 .
\end{split}
\end{align}
 We solve this equation for $x_0$ to get  \begin{align}
 y_0 &= \frac{\qty(\eta + \sigma^2\bar{\chi})\qty(\sum_jA_{j0}x_{j \setminus 0} + b_0) + z^*\qty(\sum_jA_{0j}y_{j \setminus 0} + a_0)}{|z|^2 +(\eta  + \sigma^2\bar{\chi} ) ^2}
 \end{align}
 By definition of the susceptibilities and self-consistency of the mean field cavity equations, we know that
 \begin{align}
 \bar{\chi} &= \expval{\pdv{y_0}{b_0}} = \frac{  (\eta +\sigma^2\bar{\chi})}{|z|^2 +(\eta  +  \sigma^2\bar{\chi} ) ^2}, \label{Eq:chicircle}
 \end{align}
 and
 \begin{align}
  \bar{\nu}^{(y)} &= \expval{\pdv{y_0}{a_0}} = \frac{z^*}{|z|^2 +(\eta  + \sigma^2\bar{\chi} ) ^2}.  \label{Eq:nucircle}
 \end{align} 

Next, we examine the special case where $\eta=0$.
 We further simplify this expression by exploiting the fact that Eq.~ \eqref{Eq:chicircle} gives a cubic self-consistency equation for $\chi$. 
  When $\eta=0$, we get
 \be
 \sigma^4 \bar{\chi}^3  + \bar{\chi}(|z|^2-\sigma^2) =0.\label{Eq:circledomain}
 \ee
which  has two solutions:
\begin{align}
\bar{\chi}=0 \qor \sigma^4\bar{\chi}^2 +|z|^2 =\sigma^2.
\end{align}
Substituting the first solution $\bar{\chi}=0$ into the formula for the spectral density gives
\begin{multline}
 \rho_A(u,v) ={1 \over \pi} \eval{\partial_{z^*}\bar{\nu}^{(y)}}_{\eta=0} ={1 \over \pi} \partial_{z^*} {z^* \over  |z|^2 + \sigma^4\bar{\chi}^2  } \\
= {1 \over \pi} \partial_{z^*} {z^* \over  |z|^2} = {1 \over \pi} \partial_{z^*} {1 \over  z} = 0
\label{eq:rhocircle}
\end{multline}
while the second solution yields
\begin{align}
 \rho_A(u,v) = {1 \over \pi} \partial_{z^*} {z^* \over  |z|^2 +  \sigma^4\bar{\chi}^2} = {1 \over \pi} \partial_{z^*} {z^* \over  \sigma^2} = \frac{1}{\pi\sigma^2}.
\end{align}
All together, we find
\begin{align}
\rho_A(u,v) &= \left\{\begin{array}{cl}
{1 \over  \pi \sigma^2} & \sigma^4\bar{\chi}^2 +|z|^2 =\sigma^2 \\
     0 & \qif \bar{\chi}=0.
\end{array}\right.
\end{align}
 We would like translate this into a condition on $z$ not $\bar{\chi}$. To do so, we note that  
$\sigma^4\bar{\chi}^2  +|z|^2 =\sigma^2$ implies that $|z|^2 \le |\sigma|^2$. Since the two solutions must match at $\chi=0$, this
implies that the density takes the form 
\begin{align}
\rho_A(u,v) &= \left\{\begin{array}{cl}
{1 \over  \pi \sigma^2} & \qif  |z|^2 \le \sigma^2  \\
     0 & \qif |z|^2 \ge \sigma^2,
\end{array}\right.
\end{align}
 which is precisely the circle law.
 
 \subsection{Elliptic Law}
 
 We now generalize this basic calculation to derive the Ellispe law for correlated matrices where $\zeta \neq 0$. We start once again
 with Eq. \ref{Eq:cavity-full-ellipse} and solve for $y_0$ to get
 \begin{widetext}
  \begin{align}
 y_0 &= \frac{(z^*+\zeta\sigma^2\bar{\nu})\qty(\sum_jA_{0j}y_{j \setminus 0} + a_0) + \qty(\eta + \sigma^2\bar{\chi}) \qty(\sum_jA_{j0}x_{j \setminus 0} + b_0)}{|(z+\zeta\sigma^2 \bar{\nu}^* )|^2+ (\eta  +\sigma^2 \bar{\chi} )^2}.
 \end{align}
 We then use the usual cavity arguments to write
 \begin{align}
 \bar{\chi} &= \expval{\pdv{y_0}{b_0}} = \frac{  (\eta + \sigma^2\bar{\chi})}{|(z+ \zeta\sigma^2 \bar{\nu}^* )|^2 + (\eta  + \sigma^2 \bar{\chi} )^2}\\
  \bar{\nu}^{(y)} &= \expval{\pdv{y_0}{a_0}} = \frac{z^*+\zeta\sigma^2\bar{\nu}}{|(z+ \zeta\sigma^2 \bar{\nu}^* )|^2 + (\eta  +\sigma^2 \bar{\chi} )^2} .
 \end{align}
 
To calculate the density, once again we will make use of  Eq.~\eqref{Eq:density-complex}. For $\eta=0$, notice the equation for $\bar{\chi}$ has two 
solutions 
\begin{align}
\bar{\chi}=0 \qor  |(z- \zeta\sigma^2 \bar{\nu}^* )|^2 + \sigma^4 \bar{\chi}^2=\sigma^2.
\end{align}
Plugging this into the expressions above yields (using the identity $\bar{\nu}^*=-\bar{\nu}^{(y)}$)
\begin{align}
  \bar{\nu}^{(y)} &= \left\{\begin{array}{cl}
{1 \over (z- \zeta\sigma^2 \bar{\nu}^{(y)} ) } & \qif  \chi=0  \\
 \frac{z^*-\zeta\sigma^2(\bar{\nu}^{(y)})^*}{\sigma^2}    & \qif  |z- \zeta\sigma^2 \bar{\nu}^{(y)} |^2 + \sigma^4 \bar{\chi}^2=\sigma^2,
\end{array}\right.
\end{align}
\end{widetext}
When $\chi=0$, the top expression in the equation above implies that $\bar{\nu}^{(y)}$ is an analytic function of $z$ and does
not depend on $z^*$. Hence, from Eq.~\eqref{Eq:density-complex} we conclude
that when $\chi=0$ that $\rho_A(u,v)=0$. When $|z- \zeta\sigma^2 \bar{\nu}^{(y)} |^2 + \sigma^4 \bar{\chi}^2=\sigma^2$,
we can easily solve for $\bar{\nu}^{(y)}$ by combining the bottom equation with its complex conjugate to get
\be
 \bar{\nu}^{(y)}= \frac{z^*-\zeta z}{\sigma^2 (1 -\zeta^2)}
\ee
Using Eq.~\eqref{Eq:density-complex}, we conclude that for this solution the density is then given by
 $\rho_A(x,y) =  {1  \over  \pi \sigma^2 (1-\zeta^2)}$. Summarizing, we find that
 \begin{align}
\rho_A(u,v) &= \left\{\begin{array}{cl}
{1 \over \sigma^2 \pi (1-\zeta^2)} & \qif  |z - \zeta\sigma^2 \bar{\nu}^{(y)} |^2 + \sigma^4 \bar{\chi}^2=\sigma^2 \\\
     0 & \qotherwise
\end{array}\right.
\end{align}
Notice that $|z - \zeta\sigma^2 \bar{\nu}^{(y)} |^2 + \sigma^4 \bar{\chi}^2=\sigma^2$ implies that $|z - \zeta\sigma^2 \bar{\nu}^{(y)} |^2 \le \sigma^2$.
Plugging in the explicit expression for $ \bar{\nu}^{(y)}$ into this  inequality and writing $z=u+iv$ allows us to rewrite this inequality in terms of $u$ and
$v$ as 
\be
{ u^2 \over (1+ \zeta)^2}  +{ v^2 \over (1-\zeta)^2  } \le \sigma^2
\ee
This is the precisely the elliptic law first derived by Ginbre and Girko \cite{girko1986elliptic}. 
To summarize, for a general $\zeta$, the spectral
density is given by
 \begin{align}
\rho_A(u,v) &= \left\{\begin{array}{cl}
{1 \over \sigma^2 \pi (1-\zeta^2)} & \qif  { u^2 \over (1+ \zeta)^2}  +{ v^2 \over (1-\zeta)^2  } \le \sigma^2 \\\
     0 & \qotherwise
\end{array}\right.
\end{align}
When $\zeta=0$, these reduces to the circle law derived above.

\section{Conclusion}

In this paper we present a simple, perturbative approach for calculating spectral densities for a variety of random matrix ensembles
in the thermodynamic limit we call the Perturbative Resolvent Method (PRM). The key idea of the PRM is to construct a system of random
linear equations and ask how the solution to these equations change in response to slight perturbations. The key mathematical
quantity that we make use of are appropriately defined susceptibilities which we show are directly related to Green's functions that can be
used to calculate the spectral densities. 

We have illustrated the generality and power of the method by providing simple derivations of the 
Wigner Semi-circle Law for symmetric matrices, the Marchenko-Pastur Law for Wishart matrices, the spectral density of a simple product Wishart matrix,
 and the Circle and elliptic laws
for real random matrices. In general, we expect the PRM will be useful for relating approaches based on the zero-temperature cavity method to 
results in RMT. For example, we have found the PRM can offer a new perspective for understanding problems in
ecology \cite{advani2018statistical, cui2020effect, cui2019diverse}, machine learning 
\cite{rocks2020memorizing},  and statistical inference problems such as compressed sensing \cite{ramezanali2015cavity} where zero-temperature
cavity calculations have been used to identify phase transitions. More generally, we hope that the PRM can serve as a useful method for analyzing 
disordered systems and random matrices.

\section{Acknowledgments}
We are extremely grateful to Robert Marsland III  for many useful conversations. 
This work was funded by a Simons Investigator in MMLS award and NIH NIGMS R35GM119461 grant to PM.

\bibliography{rmtcavity.bib}

\begin{thebibliography}{32}%
\makeatletter
\providecommand \@ifxundefined [1]{%
 \@ifx{#1\undefined}
}%
\providecommand \@ifnum [1]{%
 \ifnum #1\expandafter \@firstoftwo
 \else \expandafter \@secondoftwo
 \fi
}%
\providecommand \@ifx [1]{%
 \ifx #1\expandafter \@firstoftwo
 \else \expandafter \@secondoftwo
 \fi
}%
\providecommand \natexlab [1]{#1}%
\providecommand \enquote  [1]{``#1''}%
\providecommand \bibnamefont  [1]{#1}%
\providecommand \bibfnamefont [1]{#1}%
\providecommand \citenamefont [1]{#1}%
\providecommand \href@noop [0]{\@secondoftwo}%
\providecommand \href [0]{\begingroup \@sanitize@url \@href}%
\providecommand \@href[1]{\@@startlink{#1}\@@href}%
\providecommand \@@href[1]{\endgroup#1\@@endlink}%
\providecommand \@sanitize@url [0]{\catcode `\\12\catcode `\$12\catcode
  `\&12\catcode `\#12\catcode `\^12\catcode `\_12\catcode `\%12\relax}%
\providecommand \@@startlink[1]{}%
\providecommand \@@endlink[0]{}%
\providecommand \url  [0]{\begingroup\@sanitize@url \@url }%
\providecommand \@url [1]{\endgroup\@href {#1}{\urlprefix }}%
\providecommand \urlprefix  [0]{URL }%
\providecommand \Eprint [0]{\href }%
\providecommand \doibase [0]{http://dx.doi.org/}%
\providecommand \selectlanguage [0]{\@gobble}%
\providecommand \bibinfo  [0]{\@secondoftwo}%
\providecommand \bibfield  [0]{\@secondoftwo}%
\providecommand \translation [1]{[#1]}%
\providecommand \BibitemOpen [0]{}%
\providecommand \bibitemStop [0]{}%
\providecommand \bibitemNoStop [0]{.\EOS\space}%
\providecommand \EOS [0]{\spacefactor3000\relax}%
\providecommand \BibitemShut  [1]{\csname bibitem#1\endcsname}%
\let\auto@bib@innerbib\@empty
\bibitem [{\citenamefont {Dyson}(1962)}]{dyson1962statistical}%
  \BibitemOpen
  \bibfield  {author} {\bibinfo {author} {\bibfnamefont {F.~J.}\ \bibnamefont
  {Dyson}},\ }\href@noop {} {\bibfield  {journal} {\bibinfo  {journal} {Journal
  of Mathematical Physics}\ }\textbf {\bibinfo {volume} {3}},\ \bibinfo {pages}
  {140} (\bibinfo {year} {1962})}\BibitemShut {NoStop}%
\bibitem [{\citenamefont {Auffinger}\ \emph {et~al.}(2013)\citenamefont
  {Auffinger}, \citenamefont {Arous},\ and\ \citenamefont
  {{\v{C}}ern{\`y}}}]{auffinger2013random}%
  \BibitemOpen
  \bibfield  {author} {\bibinfo {author} {\bibfnamefont {A.}~\bibnamefont
  {Auffinger}}, \bibinfo {author} {\bibfnamefont {G.~B.}\ \bibnamefont
  {Arous}}, \ and\ \bibinfo {author} {\bibfnamefont {J.}~\bibnamefont
  {{\v{C}}ern{\`y}}},\ }\href@noop {} {\bibfield  {journal} {\bibinfo
  {journal} {Communications on Pure and Applied Mathematics}\ }\textbf
  {\bibinfo {volume} {66}},\ \bibinfo {pages} {165} (\bibinfo {year}
  {2013})}\BibitemShut {NoStop}%
\bibitem [{\citenamefont {Kriecherbauer}\ \emph {et~al.}(2001)\citenamefont
  {Kriecherbauer}, \citenamefont {Marklof},\ and\ \citenamefont
  {Soshnikov}}]{kriecherbauer2001random}%
  \BibitemOpen
  \bibfield  {author} {\bibinfo {author} {\bibfnamefont {T.}~\bibnamefont
  {Kriecherbauer}}, \bibinfo {author} {\bibfnamefont {J.}~\bibnamefont
  {Marklof}}, \ and\ \bibinfo {author} {\bibfnamefont {A.}~\bibnamefont
  {Soshnikov}},\ }\href@noop {} {\bibfield  {journal} {\bibinfo  {journal}
  {Proceedings of the National Academy of Sciences}\ }\textbf {\bibinfo
  {volume} {98}},\ \bibinfo {pages} {10531} (\bibinfo {year}
  {2001})}\BibitemShut {NoStop}%
\bibitem [{\citenamefont {May}(1972)}]{may1972will}%
  \BibitemOpen
  \bibfield  {author} {\bibinfo {author} {\bibfnamefont {R.~M.}\ \bibnamefont
  {May}},\ }\href@noop {} {\bibfield  {journal} {\bibinfo  {journal} {Nature}\
  }\textbf {\bibinfo {volume} {238}},\ \bibinfo {pages} {413} (\bibinfo {year}
  {1972})}\BibitemShut {NoStop}%
\bibitem [{\citenamefont {Allesina}\ and\ \citenamefont
  {Tang}(2015)}]{allesina2015stability}%
  \BibitemOpen
  \bibfield  {author} {\bibinfo {author} {\bibfnamefont {S.}~\bibnamefont
  {Allesina}}\ and\ \bibinfo {author} {\bibfnamefont {S.}~\bibnamefont
  {Tang}},\ }\href@noop {} {\bibfield  {journal} {\bibinfo  {journal}
  {Population Ecology}\ }\textbf {\bibinfo {volume} {57}},\ \bibinfo {pages}
  {63} (\bibinfo {year} {2015})}\BibitemShut {NoStop}%
\bibitem [{\citenamefont {Biroli}\ \emph {et~al.}(2018)\citenamefont {Biroli},
  \citenamefont {Bunin},\ and\ \citenamefont
  {Cammarota}}]{biroli2018marginally}%
  \BibitemOpen
  \bibfield  {author} {\bibinfo {author} {\bibfnamefont {G.}~\bibnamefont
  {Biroli}}, \bibinfo {author} {\bibfnamefont {G.}~\bibnamefont {Bunin}}, \
  and\ \bibinfo {author} {\bibfnamefont {C.}~\bibnamefont {Cammarota}},\
  }\href@noop {} {\bibfield  {journal} {\bibinfo  {journal} {New Journal of
  Physics}\ }\textbf {\bibinfo {volume} {20}},\ \bibinfo {pages} {083051}
  (\bibinfo {year} {2018})}\BibitemShut {NoStop}%
\bibitem [{\citenamefont {Couillet}\ and\ \citenamefont
  {Debbah}(2011)}]{couillet2011random}%
  \BibitemOpen
  \bibfield  {author} {\bibinfo {author} {\bibfnamefont {R.}~\bibnamefont
  {Couillet}}\ and\ \bibinfo {author} {\bibfnamefont {M.}~\bibnamefont
  {Debbah}},\ }\href@noop {} {\emph {\bibinfo {title} {Random matrix methods
  for wireless communications}}}\ (\bibinfo  {publisher} {Cambridge University
  Press},\ \bibinfo {year} {2011})\BibitemShut {NoStop}%
\bibitem [{\citenamefont {Livan}\ \emph {et~al.}(2018)\citenamefont {Livan},
  \citenamefont {Novaes},\ and\ \citenamefont {Vivo}}]{livan2018introduction}%
  \BibitemOpen
  \bibfield  {author} {\bibinfo {author} {\bibfnamefont {G.}~\bibnamefont
  {Livan}}, \bibinfo {author} {\bibfnamefont {M.}~\bibnamefont {Novaes}}, \
  and\ \bibinfo {author} {\bibfnamefont {P.}~\bibnamefont {Vivo}},\ }\href@noop
  {} {\emph {\bibinfo {title} {Introduction to random matrices: theory and
  practice}}},\ Vol.~\bibinfo {volume} {26}\ (\bibinfo  {publisher}
  {Springer},\ \bibinfo {year} {2018})\BibitemShut {NoStop}%
\bibitem [{\citenamefont {K{\"u}hn}(2008)}]{kuhn2008spectra}%
  \BibitemOpen
  \bibfield  {author} {\bibinfo {author} {\bibfnamefont {R.}~\bibnamefont
  {K{\"u}hn}},\ }\href@noop {} {\bibfield  {journal} {\bibinfo  {journal}
  {Journal of Physics A: Mathematical and Theoretical}\ }\textbf {\bibinfo
  {volume} {41}},\ \bibinfo {pages} {295002} (\bibinfo {year}
  {2008})}\BibitemShut {NoStop}%
\bibitem [{\citenamefont {Sengupta}\ and\ \citenamefont
  {Mitra}(1999)}]{sengupta1999distributions}%
  \BibitemOpen
  \bibfield  {author} {\bibinfo {author} {\bibfnamefont {A.~M.}\ \bibnamefont
  {Sengupta}}\ and\ \bibinfo {author} {\bibfnamefont {P.~P.}\ \bibnamefont
  {Mitra}},\ }\href@noop {} {\bibfield  {journal} {\bibinfo  {journal}
  {Physical Review E}\ }\textbf {\bibinfo {volume} {60}},\ \bibinfo {pages}
  {3389} (\bibinfo {year} {1999})}\BibitemShut {NoStop}%
\bibitem [{\citenamefont {Rogers}\ \emph {et~al.}(2008)\citenamefont {Rogers},
  \citenamefont {Castillo}, \citenamefont {K{\"u}hn},\ and\ \citenamefont
  {Takeda}}]{rogers2008cavity}%
  \BibitemOpen
  \bibfield  {author} {\bibinfo {author} {\bibfnamefont {T.}~\bibnamefont
  {Rogers}}, \bibinfo {author} {\bibfnamefont {I.~P.}\ \bibnamefont
  {Castillo}}, \bibinfo {author} {\bibfnamefont {R.}~\bibnamefont {K{\"u}hn}},
  \ and\ \bibinfo {author} {\bibfnamefont {K.}~\bibnamefont {Takeda}},\
  }\href@noop {} {\bibfield  {journal} {\bibinfo  {journal} {Physical Review
  E}\ }\textbf {\bibinfo {volume} {78}},\ \bibinfo {pages} {031116} (\bibinfo
  {year} {2008})}\BibitemShut {NoStop}%
\bibitem [{\citenamefont {Rogers}\ and\ \citenamefont
  {Castillo}(2009)}]{rogers2009cavity}%
  \BibitemOpen
  \bibfield  {author} {\bibinfo {author} {\bibfnamefont {T.}~\bibnamefont
  {Rogers}}\ and\ \bibinfo {author} {\bibfnamefont {I.~P.}\ \bibnamefont
  {Castillo}},\ }\href@noop {} {\bibfield  {journal} {\bibinfo  {journal}
  {Physical Review E}\ }\textbf {\bibinfo {volume} {79}},\ \bibinfo {pages}
  {012101} (\bibinfo {year} {2009})}\BibitemShut {NoStop}%
\bibitem [{\citenamefont {Advani}\ \emph {et~al.}(2018)\citenamefont {Advani},
  \citenamefont {Bunin},\ and\ \citenamefont {Mehta}}]{advani2018statistical}%
  \BibitemOpen
  \bibfield  {author} {\bibinfo {author} {\bibfnamefont {M.}~\bibnamefont
  {Advani}}, \bibinfo {author} {\bibfnamefont {G.}~\bibnamefont {Bunin}}, \
  and\ \bibinfo {author} {\bibfnamefont {P.}~\bibnamefont {Mehta}},\
  }\href@noop {} {\bibfield  {journal} {\bibinfo  {journal} {Journal of
  Statistical Mechanics: Theory and Experiment}\ }\textbf {\bibinfo {volume}
  {2018}},\ \bibinfo {pages} {033406} (\bibinfo {year} {2018})}\BibitemShut
  {NoStop}%
\bibitem [{\citenamefont {Mehta}\ \emph {et~al.}(2018)\citenamefont {Mehta},
  \citenamefont {Cui}, \citenamefont {Wang},\ and\ \citenamefont
  {Marsland~III}}]{mehta2018constrained}%
  \BibitemOpen
  \bibfield  {author} {\bibinfo {author} {\bibfnamefont {P.}~\bibnamefont
  {Mehta}}, \bibinfo {author} {\bibfnamefont {W.}~\bibnamefont {Cui}}, \bibinfo
  {author} {\bibfnamefont {C.-H.}\ \bibnamefont {Wang}}, \ and\ \bibinfo
  {author} {\bibfnamefont {R.}~\bibnamefont {Marsland~III}},\ }\href@noop {}
  {\bibfield  {journal} {\bibinfo  {journal} {Physical Review E}\ }\textbf
  {\bibinfo {volume} {99}},\ \bibinfo {pages} {052111} (\bibinfo {year}
  {2018})}\BibitemShut {NoStop}%
\bibitem [{\citenamefont {Cui}\ \emph {et~al.}(2020)\citenamefont {Cui},
  \citenamefont {Marsland~III},\ and\ \citenamefont {Mehta}}]{cui2020effect}%
  \BibitemOpen
  \bibfield  {author} {\bibinfo {author} {\bibfnamefont {W.}~\bibnamefont
  {Cui}}, \bibinfo {author} {\bibfnamefont {R.}~\bibnamefont {Marsland~III}}, \
  and\ \bibinfo {author} {\bibfnamefont {P.}~\bibnamefont {Mehta}},\
  }\href@noop {} {\bibfield  {journal} {\bibinfo  {journal} {Physical Review
  Letters}\ }\textbf {\bibinfo {volume} {125}},\ \bibinfo {pages} {048101}
  (\bibinfo {year} {2020})}\BibitemShut {NoStop}%
\bibitem [{\citenamefont {Cui}\ \emph {et~al.}(2019)\citenamefont {Cui},
  \citenamefont {Marsland~III},\ and\ \citenamefont {Mehta}}]{cui2019diverse}%
  \BibitemOpen
  \bibfield  {author} {\bibinfo {author} {\bibfnamefont {W.}~\bibnamefont
  {Cui}}, \bibinfo {author} {\bibfnamefont {R.}~\bibnamefont {Marsland~III}}, \
  and\ \bibinfo {author} {\bibfnamefont {P.}~\bibnamefont {Mehta}},\
  }\href@noop {} {\bibfield  {journal} {\bibinfo  {journal} {arXiv preprint
  arXiv:1904.02610}\ } (\bibinfo {year} {2019})}\BibitemShut {NoStop}%
\bibitem [{\citenamefont {Feinberg}\ and\ \citenamefont
  {Zee}(1997{\natexlab{a}})}]{feinberg1997non}%
  \BibitemOpen
  \bibfield  {author} {\bibinfo {author} {\bibfnamefont {J.}~\bibnamefont
  {Feinberg}}\ and\ \bibinfo {author} {\bibfnamefont {A.}~\bibnamefont {Zee}},\
  }\href@noop {} {\bibfield  {journal} {\bibinfo  {journal} {Nuclear Physics
  B}\ }\textbf {\bibinfo {volume} {504}},\ \bibinfo {pages} {579} (\bibinfo
  {year} {1997}{\natexlab{a}})}\BibitemShut {NoStop}%
\bibitem [{\citenamefont {Feinberg}\ and\ \citenamefont
  {Zee}(1997{\natexlab{b}})}]{feinberg1997non2}%
  \BibitemOpen
  \bibfield  {author} {\bibinfo {author} {\bibfnamefont {J.}~\bibnamefont
  {Feinberg}}\ and\ \bibinfo {author} {\bibfnamefont {A.}~\bibnamefont {Zee}},\
  }\href@noop {} {\bibfield  {journal} {\bibinfo  {journal} {Nuclear Physics
  B}\ }\textbf {\bibinfo {volume} {501}},\ \bibinfo {pages} {643} (\bibinfo
  {year} {1997}{\natexlab{b}})}\BibitemShut {NoStop}%
\bibitem [{\citenamefont {Br{\'e}zin}\ and\ \citenamefont
  {Zee}(1995)}]{brezin1995universal}%
  \BibitemOpen
  \bibfield  {author} {\bibinfo {author} {\bibfnamefont {E.}~\bibnamefont
  {Br{\'e}zin}}\ and\ \bibinfo {author} {\bibfnamefont {A.}~\bibnamefont
  {Zee}},\ }\href@noop {} {\bibfield  {journal} {\bibinfo  {journal} {Nuclear
  Physics B}\ }\textbf {\bibinfo {volume} {453}},\ \bibinfo {pages} {531}
  (\bibinfo {year} {1995})}\BibitemShut {NoStop}%
\bibitem [{\citenamefont {Wigner}(1993)}]{wigner1993characteristic}%
  \BibitemOpen
  \bibfield  {author} {\bibinfo {author} {\bibfnamefont {E.~P.}\ \bibnamefont
  {Wigner}},\ }in\ \href@noop {} {\emph {\bibinfo {booktitle} {The Collected
  Works of Eugene Paul Wigner}}}\ (\bibinfo  {publisher} {Springer},\ \bibinfo
  {year} {1993})\ pp.\ \bibinfo {pages} {524--540}\BibitemShut {NoStop}%
\bibitem [{\citenamefont {Marchenko}\ and\ \citenamefont
  {Pastur}(1967)}]{marchenko1967distribution}%
  \BibitemOpen
  \bibfield  {author} {\bibinfo {author} {\bibfnamefont {V.~A.}\ \bibnamefont
  {Marchenko}}\ and\ \bibinfo {author} {\bibfnamefont {L.~A.}\ \bibnamefont
  {Pastur}},\ }\href@noop {} {\bibfield  {journal} {\bibinfo  {journal}
  {Matematicheskii Sbornik}\ }\textbf {\bibinfo {volume} {114}},\ \bibinfo
  {pages} {507} (\bibinfo {year} {1967})}\BibitemShut {NoStop}%
\bibitem [{\citenamefont {Girko}(1985)}]{girko1985circular}%
  \BibitemOpen
  \bibfield  {author} {\bibinfo {author} {\bibfnamefont {V.~L.}\ \bibnamefont
  {Girko}},\ }\href@noop {} {\bibfield  {journal} {\bibinfo  {journal} {Theory
  of Probability \& Its Applications}\ }\textbf {\bibinfo {volume} {29}},\
  \bibinfo {pages} {694} (\bibinfo {year} {1985})}\BibitemShut {NoStop}%
\bibitem [{\citenamefont {Girko}(1986)}]{girko1986elliptic}%
  \BibitemOpen
  \bibfield  {author} {\bibinfo {author} {\bibfnamefont {V.}~\bibnamefont
  {Girko}},\ }\href@noop {} {\bibfield  {journal} {\bibinfo  {journal} {Theory
  of Probability \& Its Applications}\ }\textbf {\bibinfo {volume} {30}},\
  \bibinfo {pages} {677} (\bibinfo {year} {1986})}\BibitemShut {NoStop}%
\bibitem [{\citenamefont {Rogers}(2010)}]{rogers2010new}%
  \BibitemOpen
  \bibfield  {author} {\bibinfo {author} {\bibfnamefont {T.}~\bibnamefont
  {Rogers}},\ }\emph {\bibinfo {title} {New results on the spectral density of
  random matrices}},\ \href@noop {} {Ph.D. thesis},\ \bibinfo  {school} {King's
  College London} (\bibinfo {year} {2010})\BibitemShut {NoStop}%
\bibitem [{\citenamefont {Tao}(2012)}]{tao2012topics}%
  \BibitemOpen
  \bibfield  {author} {\bibinfo {author} {\bibfnamefont {T.}~\bibnamefont
  {Tao}},\ }\href@noop {} {\emph {\bibinfo {title} {Topics in random matrix
  theory}}},\ Vol.\ \bibinfo {volume} {132}\ (\bibinfo  {publisher} {American
  Mathematical Soc.},\ \bibinfo {year} {2012})\BibitemShut {NoStop}%
\bibitem [{\citenamefont {Mingo}\ and\ \citenamefont
  {Speicher}(2017)}]{mingo2017free}%
  \BibitemOpen
  \bibfield  {author} {\bibinfo {author} {\bibfnamefont {J.~A.}\ \bibnamefont
  {Mingo}}\ and\ \bibinfo {author} {\bibfnamefont {R.}~\bibnamefont
  {Speicher}},\ }\href@noop {} {\emph {\bibinfo {title} {Free probability and
  random matrices}}},\ Vol.~\bibinfo {volume} {35}\ (\bibinfo  {publisher}
  {Springer},\ \bibinfo {year} {2017})\BibitemShut {NoStop}%
\bibitem [{\citenamefont {Bai}\ and\ \citenamefont
  {Silverstein}(2010)}]{bai2010spectral}%
  \BibitemOpen
  \bibfield  {author} {\bibinfo {author} {\bibfnamefont {Z.}~\bibnamefont
  {Bai}}\ and\ \bibinfo {author} {\bibfnamefont {J.~W.}\ \bibnamefont
  {Silverstein}},\ }\href@noop {} {\emph {\bibinfo {title} {Spectral analysis
  of large dimensional random matrices}}},\ Vol.~\bibinfo {volume} {20}\
  (\bibinfo  {publisher} {Springer},\ \bibinfo {year} {2010})\BibitemShut
  {NoStop}%
\bibitem [{\citenamefont {Lu}\ and\ \citenamefont
  {Shiou}(2002)}]{lu2002inverses}%
  \BibitemOpen
  \bibfield  {author} {\bibinfo {author} {\bibfnamefont {T.-T.}\ \bibnamefont
  {Lu}}\ and\ \bibinfo {author} {\bibfnamefont {S.-H.}\ \bibnamefont {Shiou}},\
  }\href@noop {} {\bibfield  {journal} {\bibinfo  {journal} {Computers \&
  Mathematics with Applications}\ }\textbf {\bibinfo {volume} {43}},\ \bibinfo
  {pages} {119} (\bibinfo {year} {2002})}\BibitemShut {NoStop}%
\bibitem [{\citenamefont {Burda}\ \emph {et~al.}(2010)\citenamefont {Burda},
  \citenamefont {Jarosz}, \citenamefont {Livan}, \citenamefont {Nowak},\ and\
  \citenamefont {Swiech}}]{burda2010eigenvalues}%
  \BibitemOpen
  \bibfield  {author} {\bibinfo {author} {\bibfnamefont {Z.}~\bibnamefont
  {Burda}}, \bibinfo {author} {\bibfnamefont {A.}~\bibnamefont {Jarosz}},
  \bibinfo {author} {\bibfnamefont {G.}~\bibnamefont {Livan}}, \bibinfo
  {author} {\bibfnamefont {M.~A.}\ \bibnamefont {Nowak}}, \ and\ \bibinfo
  {author} {\bibfnamefont {A.}~\bibnamefont {Swiech}},\ }\href@noop {}
  {\bibfield  {journal} {\bibinfo  {journal} {Physical Review E}\ }\textbf
  {\bibinfo {volume} {82}},\ \bibinfo {pages} {061114} (\bibinfo {year}
  {2010})}\BibitemShut {NoStop}%
\bibitem [{\citenamefont {Dupic}\ and\ \citenamefont
  {Castillo}(2014)}]{dupic2014spectral}%
  \BibitemOpen
  \bibfield  {author} {\bibinfo {author} {\bibfnamefont {T.}~\bibnamefont
  {Dupic}}\ and\ \bibinfo {author} {\bibfnamefont {I.~P.}\ \bibnamefont
  {Castillo}},\ }\href@noop {} {\bibfield  {journal} {\bibinfo  {journal}
  {arXiv preprint arXiv:1401.7802}\ } (\bibinfo {year} {2014})}\BibitemShut
  {NoStop}%
\bibitem [{\citenamefont {Rocks}\ and\ \citenamefont
  {Mehta}(2020)}]{rocks2020memorizing}%
  \BibitemOpen
  \bibfield  {author} {\bibinfo {author} {\bibfnamefont {J.~W.}\ \bibnamefont
  {Rocks}}\ and\ \bibinfo {author} {\bibfnamefont {P.}~\bibnamefont {Mehta}},\
  }\href@noop {} {\bibfield  {journal} {\bibinfo  {journal} {arXiv preprint
  arXiv:2010.13933}\ } (\bibinfo {year} {2020})}\BibitemShut {NoStop}%
\bibitem [{\citenamefont {Ramezanali}\ \emph {et~al.}(2015)\citenamefont
  {Ramezanali}, \citenamefont {Mitra},\ and\ \citenamefont
  {Sengupta}}]{ramezanali2015cavity}%
  \BibitemOpen
  \bibfield  {author} {\bibinfo {author} {\bibfnamefont {M.}~\bibnamefont
  {Ramezanali}}, \bibinfo {author} {\bibfnamefont {P.~P.}\ \bibnamefont
  {Mitra}}, \ and\ \bibinfo {author} {\bibfnamefont {A.~M.}\ \bibnamefont
  {Sengupta}},\ }\href@noop {} {\bibfield  {journal} {\bibinfo  {journal}
  {arXiv preprint arXiv:1501.03194}\ } (\bibinfo {year} {2015})}\BibitemShut
  {NoStop}%
\end{thebibliography}%
\end{document}